\documentclass[journal]{IEEEtran}

\usepackage{graphicx}
\usepackage{subcaption}
\usepackage{color, colortbl}
\definecolor{Gray}{gray}{0.8}
\newcommand{\ULsubfloat}[2][\empty]
{\hbox{
  \sbox0{#2}
  \captionsetup{position=top,
  justification=centering, singlelinecheck=false}%
  \rotatebox[origin=bl]{90}{\begin{minipage}[b]{\dimexpr \ht0+\dp0}
    \subcaption{#1}
  \end{minipage}}\raisebox{\dp0}{\usebox0}%
}}

\graphicspath{{figs/}}
\begin{document}

\title{Learning Ultrasound Rendering from Cross-Sectional Model Slices for Simulated Training}

\author{Lin Zhang, Tiziano Portenier, Orcun Goksel
\thanks{L. Zhang, T. Portenier, O. Goksel are with the Computer-assisted Applications in Medicine, ETH Zurich, Switzerland.}
\thanks{O. Goksel is also with the Department of Information Technology, Uppsala University, Sweden.}
\thanks{Funding was provided by the Swiss Innovation Agency Innosuisse.}}

\maketitle

\begin{abstract}
\emph{Purpose.} Given the high level of expertise required for navigation and interpretation of ultrasound images, computational simulations can facilitate the training of such skills in virtual reality.
With ray-tracing based simulations, realistic ultrasound images can be generated.
However, due to computational constraints for interactivity, image quality typically needs to be compromised. 

\noindent \emph{Methods.} 
We propose herein to bypass any rendering and simulation process at interactive time, by conducting such simulations during a non-time-critical offline stage and then learning image translation from cross-sectional model slices to such simulated frames.
We use a generative adversarial framework with a dedicated generator architecture and input feeding scheme, which both substantially improve image quality without increase in network parameters.
Integral attenuation maps derived from cross-sectional model slices, texture-friendly strided convolutions, providing stochastic noise and input maps to intermediate layers in order to preserve locality are all shown herein to greatly facilitate such translation task.

\noindent \emph{Results.}
Given several quality metrics, the proposed method with only tissue maps as input is shown to provide comparable or superior results to a state-of-the-art that uses additional images of low-quality ultrasound renderings.
An extensive ablation study shows the need and benefits from the individual contributions utilized in this work, based on qualitative examples and quantitative ultrasound similarity metrics.
To that end, a local histogram statistics based error metric is proposed and demonstrated for visualization of local dissimilarities between ultrasound images.

\noindent \emph{Conclusion.}
A deep-learning based direct transformation from interactive tissue slices to likeness of high quality renderings allow to obviate any complex rendering process in real-time, which could enable extremely realistic ultrasound simulations on consumer-hardware by moving the time-intensive processes to a one-time, offline, preprocessing data preparation stage that can be performed on dedicated high-end hardware.

\end{abstract}

\section{Introduction}
\label{intro}
Ultrasound (US) imaging is a real-time, non-invasive and radiation-free imaging modality, making it ideal for computer-assisted interventions.
Nevertheless, the challenges in ultrasound probe manipulation and image interpretation necessitates extensive operator training.
Computer-assisted ultrasound simulation can facilitate such training in a virtual-reality environment~\cite{blum2013review,ostergaard2019simulator,ramaiah2020simulation,sun2011real}.
This does not require volunteer patients or on-site supervisors, while enabling students to learn and practice skills in a flexible, stress-free, and self-supervised manner.
Furthermore, rare-to-encounter diseases and conditions can be presented and practiced in a realistic, interactive mode.

To provide training for US probe manipulation and image interpretation, computational simulators need to correctly represent ultrasonic physical effects, e.g. view-dependent artifacts and realistic tissue texture, and operate dynamically and interactively. In the literature US simulation methods can be categorized into interpolative and generative approaches. 
The former~\cite{goksel2009b} generates 2D US images by interpolating pre-recorded 3D volumes. However, to generate a rich variety of images, a large amount of 3D volumes needs to be acquired and stored. Furthermore, it is difficult to generate novel views and contents, since the physical model is in general not considered.
In contrast to interpolative US simulation approaches, advanced generative methods~\cite{burger2013real,mattausch2018realistic,salehi2015patient,starkov2019ultrasound} allow to generate variety of images with plausible view-dependent artifacts, e.g.\ for rare pathological cases.
These techniques model ultrasonic wave propagation using ray-tracing techniques on anatomical models.
New scenes of any given anatomical model can be simulated with different imaging parameters and conditions.
There is a line of works on developing sophisticated ray tracing techniques to simulate realistic wave interactions.
A deterministic surface-based ray tracing method is introduced in~\cite{burger2013real} to simulate ultrasonic directional wave interactions.
The method is further extended in~\cite{salehi2015patient} to utilize patient-specific volumetric MRI or CT data.
A more sophisticated stochastic wave-interaction and surface modeling is proposed in~\cite{mattausch2018realistic} to overcome the simplified assumption of the deterministic ray model. 
Given a 3D anatomical model, ray-based techniques using the state-of-the-art Monte-carlo ray-tracing framework manage to simulate US images with surprisingly high realism at interactive frame rates, as shown in~\cite{mattausch2018realistic,starkov2019ultrasound} for fetal ultrasound imaging.

Recent advances in deep learning have enabled various learning based approaches for ultrasound simulation.
Generative adversarial networks~\cite{goodfellow2014generative} (GANs) are the most promising models in this regard due to their outstanding performance in generating photo-realistic images.
Conditional GANs are widely adopted for generating ultrasound images conditioned on physical input, such as calibrated physical coordinate~\cite{hu2017freehand} and echogenicity map~\cite{tom2018simulating}.
A recent publication~\cite{magnetti2020deep} has employed a generative autoencoder model learned with a large amount of tracked ultrasound data, to perform patient-specific image generation from transducer position and orientation.
To simulate realistic ultrasound speckles, the authors in~\cite{bargsten2020specklegan} have introduced a speckle layer to incorporate the physical model of speckle generation into a GAN-based data augmentation network.
In~\cite{vitale2019improving}, a cycleGAN~\cite{zhu2017unpaired} model is employed for improving the realism of simulated ultrasound images.
A constrained cycleGAN is proposed in~\cite{jafari2020cardiac} for unpaired translation from echocardiographic images acquired with point-of-care ultrasound to high-end devices.
Recently, a GAN-based image translation framework has been proposed in~\cite{zhang2020deep} for recovering high quality US images equivalent to computationally expensive ray-based simulations using low-quality and thus faster simulations, by also leveraging information from corresponding segmentation and integral attenuation maps.
However, rendering such low-quality ultrasound images still requires additional computation time and sophisticated hardware resources.

\begin{figure*}
\centering\hfill
\minipage{0.45\textwidth}
 \centering
 \includegraphics[width=0.9\linewidth]{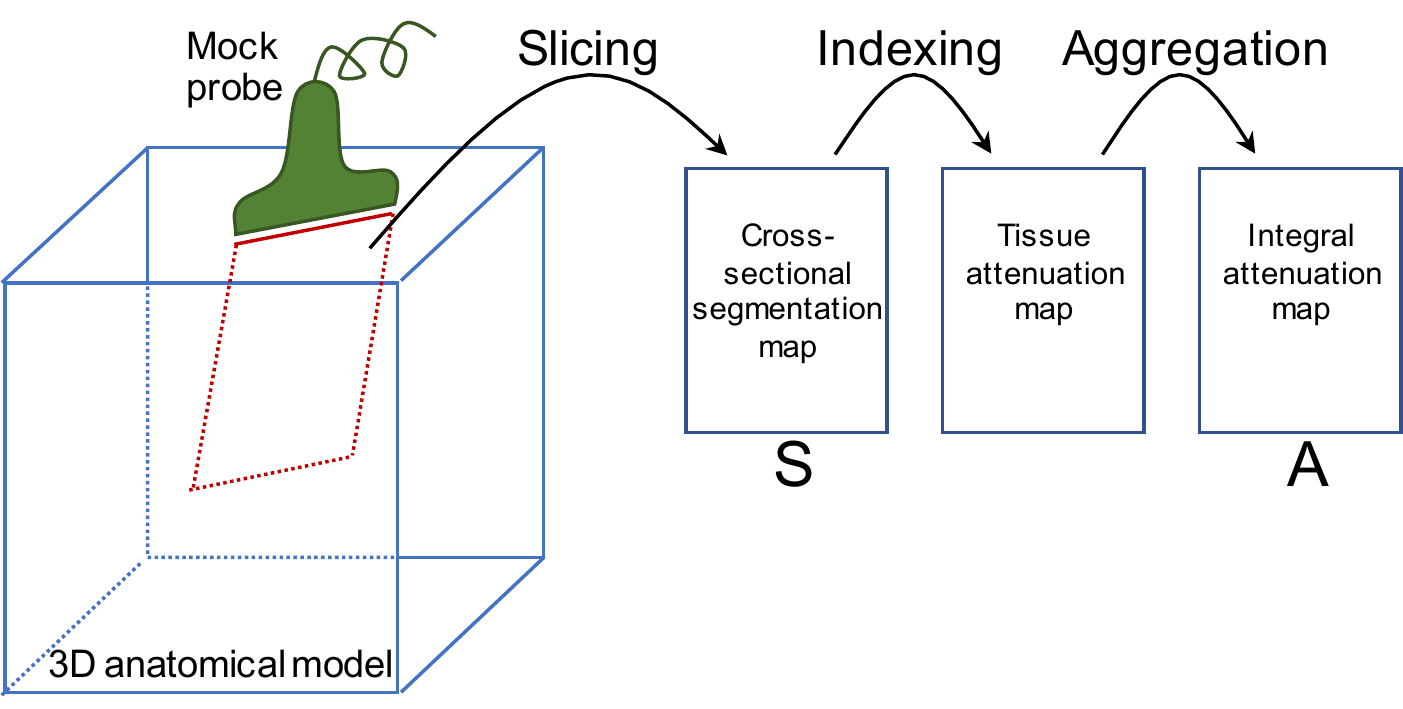}
 \caption*{(a) Per-frame generation of S \& A}
 \includegraphics[width=.6\linewidth]{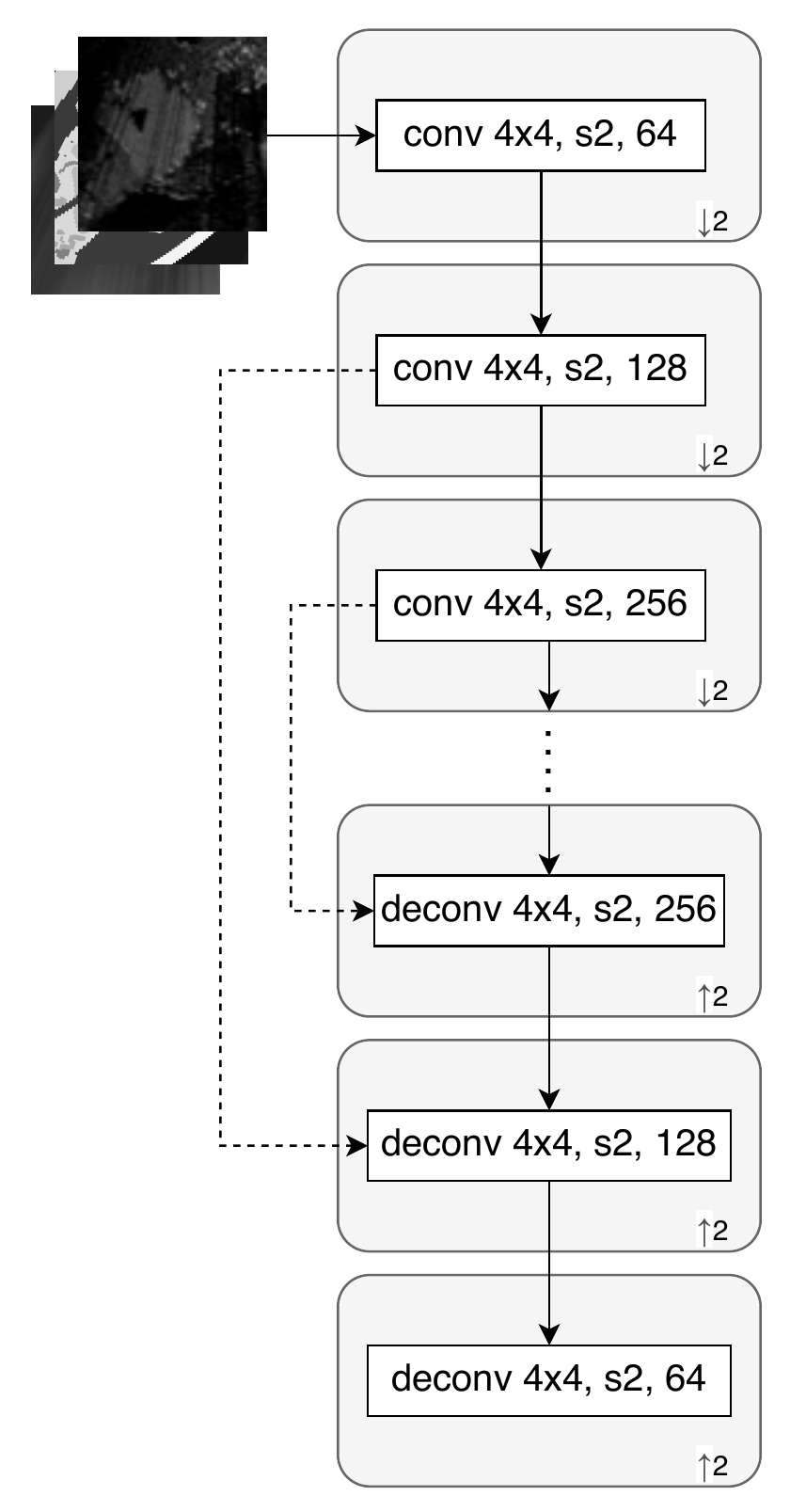}
 \caption*{(b) Generator~\cite{zhang2020deep}}
\endminipage\hfill
\minipage{0.405\textwidth}
 \centering
 \includegraphics[width=1\linewidth]{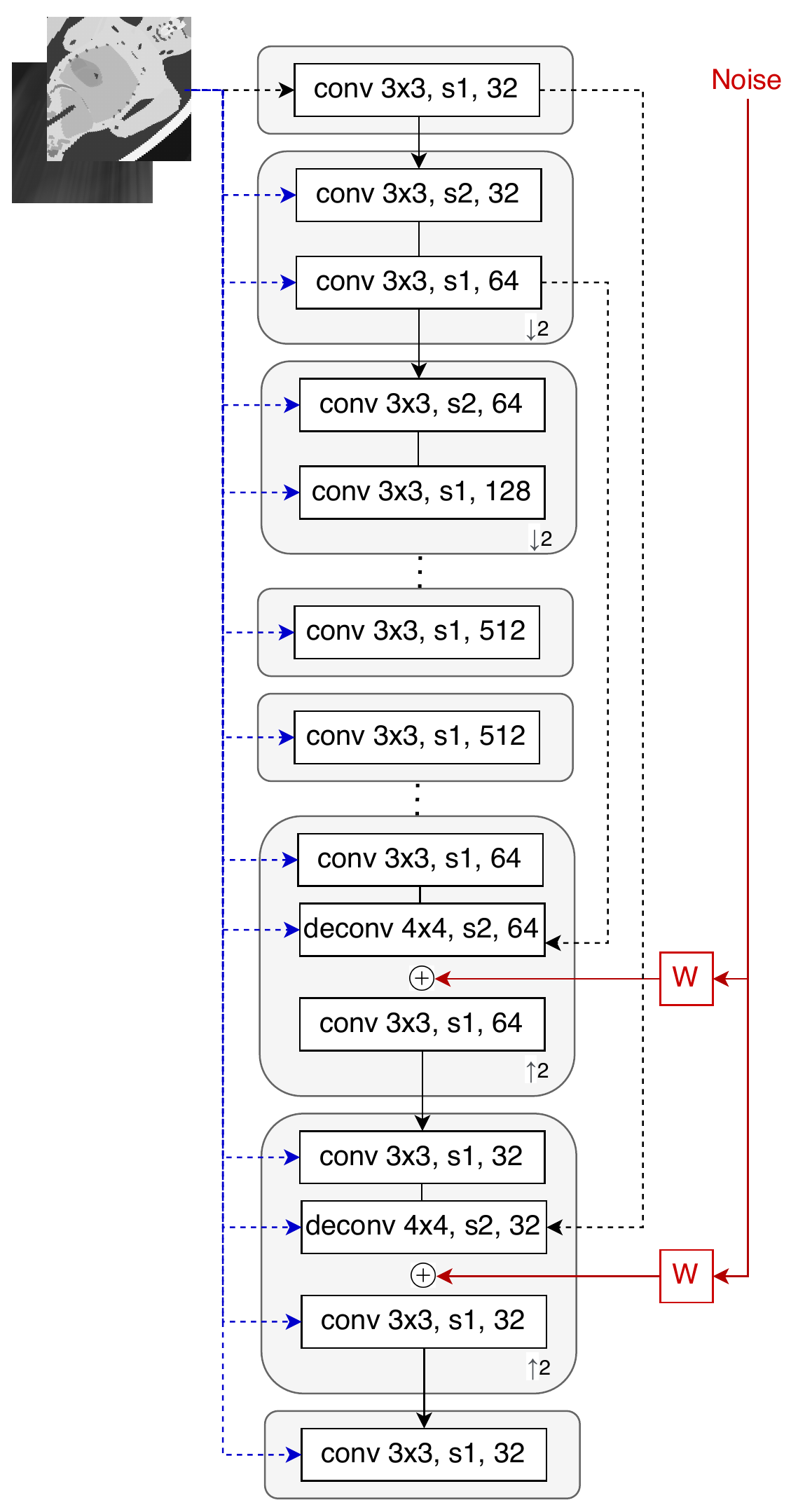}
 \vspace{-2em}\caption*{(c) Proposed generator}
\endminipage\hfill\phantom{.}
\caption{(a)~For each frame of ultrasound simulation, a cross-sectional tissue slice {\bf S} is extracted given the mock probe position and a set of 3D triangulated anatomical surfaces.
Referencing acoustic attenuation of each anatomical structure from a predefined indexed list, a spatial attenuation map is generated. Aggregating such maps along ultrasound propagation then yields the so-called integral attenuation maps {\bf A} are derived for each frame in real-time.
(b)~Network architecture of the generator used in~\cite{zhang2020deep}.
(c)~Network architecture proposed herein to address the challenges of image translation from segmentation maps to complex B-mode image content.
Here the white blocks “(de)conv $k\times k$, s$s$, $ch$” stands for a (de)convolutional layer with a filter size of $k\times k$, stride $s$, and $ch$ filters. Gray block with ↑2 or ↓2 denotes the entire up- or downsampling block with a factor of 2. Dotted black lines stand for skip connections and blue lines for concatenation of the network input. Solid red line and block represents stochastic components of the network. The term “W” denotes learned per-channel weighting.}
\label{fig:network}
\end{figure*}

In this work, we propose to learn the rendering of ultrasound images given only cross-sectional model slice / segmentation and integral attenuation maps, the latter of which can be derived from former on-the-fly and helps distill global acoustic energy information locally.
An overview of per-frame segmentation and integral attenuation map generation can be seen in Fig.~\ref{fig:network}(a).
Herein we adopt a Monte-carlo simulation presented in~\cite{mattausch2018realistic} for simulating the same abdominal scene as in previous work~\cite{zhang2020deep}.
Since low-quality images as used in~\cite{zhang2020deep} provide information about wave interactions and speckle textures, omitting this modality in the network input renders the translation task much more challenging.
We carefully inspect the generator architecture and propose several crucial architectural improvements to enable realistic quality images from merely the segmentation maps as input.
Herein we present a local feature and texture preserving generator containing
\begin{itemize}
\item connections of network input to each intermediate layer for improved information flow;
\item smaller receptive field to preserve spatial information;
\item texture-friendly strided convolutions;
\item dedicated noise images fed at each resolution.
\end{itemize}
With the above mentioned architectural changes in contrast to~\cite{zhang2020deep}, our proposed network is able to simulate US images with realistic structural and textural content.
To the best of our knowledge, this is the first work investigating deep neural networks to generate simulated ultrasound images using only structural tissue cross-sections.

\section{Methods}
\label{methods}

\subsection{Ray-based Ultrasound Simulation}
\label{simulation}
In this work we adopt a Monte-Carlo ray tracing framework presented in~\cite{mattausch2018realistic} to simulate ultrasound B-mode images for an abdominal scene.
Directional ultrasonic wave interactions such as reflection and refraction are simulated using a stochastic surface ray tracing model, which is able to create realistic looking soft shadows and tissue reflection boundaries.
Ultrasound speckle formation is modeled by convolving 3D tissue scattering representation, parametrized by a Gaussian distribution~\cite{mattausch2015scatterer}, with spatially-variant point-spread-function. 
Each tissue type is assigned a set of pre-determined acoustic properties such as acoustic impedance, attenuation, and speed-of-sound, which are modeled by hand according to literature and quality assessment by sonographers.
To obtain gray-scale B-mode images, conventional post-processing steps are applied on simulated RF data, including envelope detection, time-gain compensation, log compression, and scan conversion.
Herein we simulate the second trimester transabdominal scene of the entire abdomen. The US probe is placed at positions on a regular grid on the abdominal surface and different orientations at each position are collected to uniformly sample the fetus field-of-view.
Detailed scene information and simulation parameters are provided in Table~\ref{tab:sim_params}.
\begin{table}
\caption{Simulation parameters}
\label{tab:sim_params}
\centering
\begin{tabular}{l||r}
parameter & value \\
\hline 
Triangles fetus & 400k\\
Triangles mother & 275k\\
Image depth & 15.0 cm \\
Elevational layer & 3
\end{tabular}
\begin{tabular}{l||r}
parameter & value \\
\hline 
Transducer frequency & 8 MHz\\
Transducer field-of-view & 70$^{\circ}$\\
Axial samples & 3072 \\
Ray per scanline & 32
\end{tabular}
\end{table}
Model slice is rendered as the cross-section of the triangulated anatomical surfaces.
It provides details about the anatomical layout, thus is also referred to segmentation map herein.
The segmentation map is stored as an indexed image. Each tissue type has its unique intensity value, which is matched to a corresponding attenuation coefficient using a look-up table prior to axial aggregation for attenuation map generation.

One of the most characteristic features of US images is acoustic shadow, which is important for image interpretation, since abnormal shadows may indicate the existence of tissue modification.
This directional artifact is mainly caused by tissue attenuation, which leads to a decrease in sound wave intensity along the wave propagation path due to local tissue absorption and scattering.
Therefore, acoustic shadow is a cumulative effect which requires a large network receptive field and significant network modeling ability.
The authors in~\cite{zhang2020deep} propose to provide this cumulative information in a form of an integral attenuation map and they demonstrates that providing this additional information to the network greatly helps to reproduce acoustic shadows.
Integral attenuation maps are computed as $a=e^{-\sum_{i=0}^z \mu[i]}$ with the image depth $z$ and frequency-dependent and tissue specific attenuation coefficient $\mu$.
The acoustic shadow formation is thus approximated by accumulating the attenuation strength, quantified by $\mu$, along the wave propagation path. For each image point, the acoustic attenuation is inferred from the segmentation value using a predefined look-up table. The integral attenuation map is generated separately for each pre-scan-converted image column, which corresponds to radial US propagation direction for the given convex US probe. The integral maps are then normalized by 98 \%ile of image intensities, and scan-converted into Cartesian coordinates.

\subsection{Ultrasound Simulation using a Generative Adversarial Network}
\label{gan}
In this work, we propose a generative adversarial network (GAN) for generating ultrasound images from segmentation and integral attenuation maps.
Given segmentation maps and acoustic attenuation properties of each tissue, integral maps can be generated as in~\cite{zhang2020deep} in order to mimic wavefront traversal in tissue.
This is known from~\cite{zhang2020deep} to improve rendered shadows and acoustic amplifications by bringing global echo amplitude modifications to a local context.
In our ablation experiments we study the effect of attenuation maps, referred in the results with the acronym \emph{att}.

In contrast to~\cite{zhang2020deep}, herein we propose the following four major architectural improvements, as also illustrated in Fig.~\ref{fig:network}, which are later demonstrated to provide promising generation results in our experiments.

{\bf Input to all channels:}
To enable an effective information flow from the input images to each spatial resolution in the network, we provide the information at each intermediate layer by concatenating the maps along the channel dimension to the network activations, indicated by dotted blue connections in Fig.~\ref{fig:network}(c).
Although this was shown in~\cite{park2019semantic} to be inferior to spatially-adaptive normalization (SPADE) layers in the context of natural images, we found this approach to outperform SPADE in our application setting, since it enables the network to generate location-specific features conditioned on both segmentation and integral attenuation map.
In our ablation experiments, we refer to this concatenation approach with the acronym \emph{concat}.

{\bf Preserving spatial information:}
To further preserve the local information from input maps to output images, we use a comparatively small receptive field in our network architecture.
Since the integral attenuation map helps to transform global effects of acoustic shadows into local features, understanding and encoding the entire scene in a compact form is not required and would waste network capacity.
We therefore use 4 downsampling blocks.

{\bf Texture-friendly decoder:}
Transposed convolutions are known to be prone to characteristic checkerboard artifacts due to uneven overlap when using odd kernel sizes~\cite{odena2016deconvolution}.
Therefore they are not an ideal choice in texture generation tasks. 
To improve texture generation performance, we therefore enhance the decoding blocks by introducing additional stride-1 convolution layers, which helps to circumvent both low and high frequency artifacts.
These changes in the convolutional architecture are referred collectively using the acronym \emph{conv} in our ablation study.

{\bf Stochastic texture generation:}
To generate random speckle textures, the network needs access to a stochastic source, especially when omitting low-quality B-mode image input.
A straightforward approach to ensure a stochastic process is to feed an explicit noise image as an additional input channel. 
Recently, Karras et al.~\cite{karras2019style} proposed an alternative method, by perturbing feature channels using additive Gaussian noise with learned per-channel weighting.
This warrants the network to disentangle global and local stochastic variations.
Motivated by its astonishing performance in generating fine stochastic details such as hair, we study this approach for the generation of ultrasound textures.
Accordingly, Gaussian noise images with different sizes are generated and weighted by a learned per-channel scaling factor. 
The weighted noise channels are then added to the decoding feature layers after skip connections, shown as solid red lines in Fig.~\ref{fig:network}(c).
In the ablation study we referred to this technique with the acronym \emph{noise}.
The more conventional way of adding noise as an input layer is equivalent to replacing the low-quality image in~\cite{zhang2020deep} with noise, and therefore is referred as NSA2H in our comparative study.

We adopt the patchGAN discriminator and the training objective from~\cite{zhang2020deep}. The loss function consists of a GAN training objective $L_\mathrm{GAN}$ and an $L_1$-based data fidelity term $L_\mathrm{F}$ as follows: 
\begin{eqnarray}
    L &=& L_\mathrm{GAN}(G,D) + \lambda L_\mathrm{F} (G),\\
    L_\mathrm{GAN} &=& \mathbf{E}_{s,a,y}[\log D(y|s,a)] \nonumber\\
    &+& \mathbf{E}_{s,a}[\log (1- D(m\circ G(s,a)|s,a)],\\
    L_\mathrm{F} &=& \mathbf{E}_{s,a,y}[||y-m\circ G(s,a)||_1],
\end{eqnarray}
where $s\in S$ is segmentation map, $a\in A$ is integral attenuation map, and $m$ is the binary mask indicating the convex imaging region.
$\mathbf{E}_{s,a,y}$ and $\mathbf{E}_{s,a}$ are the expected values, respectively, over all samples of $\{s,a,y\}$ triples and $\{s,a\}$ pairs. 
The generator $G$ maps the source segmentation/attenuation map to the target US image, whereas the discriminator $D$ discriminates real and generated images conditioned on the generator input.
Before computing the loss, the generator output is pixel-wise multiplied with the binary mask m, which is denoted by the operator $\circ$.

\section{Results and Discussion}
\label{sec:2}
\paragraph{Implementation and network details.} All our models are trained using the Adam optimizer~\cite{kingma2014adam} with a learning rate of 0.0002 and exponential decay rates $\beta_1$ = 0.5 and $\beta_2$ = 0.999. 
The batch size is set to 4 and the loss weighting parameter $\lambda$ set to 100.
The leaky rectified linear unit is used in the encoder and the rectified linear unit throughout the decoder, except its output layer, which is activated using tanh.
Nonlinear activations are followed by instance normalization.
We use the same dataset as in~\cite{zhang2020deep} consisting of 6669 3-tuples (s, a, y) with an image size of $1000 \times 1386$.
The network is trained using on-the-fly cropped image patches with a size of $512\times512$ to make the training more efficient.
Randomly selected 6000 images were used for training and the rest for evaluation, following the same dataset split as in~\cite{zhang2020deep}.

\paragraph{Compared methods.} 
We refer our proposed network architecture as $SA2H$, which translates from segmentation map $S$ and integral attenuation map $A$ to a high-quality image $H$.
To evaluate individual architectural proposals of $SA2H$, we ablate each component separately and refer to as ``SA2H-component'' in the ablation studies below.
We accordingly compare $SA2H$ against the following alternatives:

\begin{itemize}
\item[$\bullet$] {\bf SA2H-att} omitting integral attenuation maps as input to the network;

\item[$\bullet$] {\bf SA2H-concat} providing segmentation and attenuation maps at the input layer only without concatenating in the hidden layers (removing the blue connections in Fig.~\ref{fig:network} (c));

\item[$\bullet$] {\bf SA2H-conv} omitting texture-friendly convolutional components in the decoder and falling back to even sized transposed convolution kernels;

\item[$\bullet$] {\bf SA2H-noise} omitting Gaussian noise images as stochastic input and provide an input noise layer instead (removing the red connections in Fig.~\ref{fig:network} (c));

\item[$\bullet$] {\bf LSA2H} the recent low-to-high-quality image translation network presented in~\cite{zhang2020deep}, which has a low-quality rendered ultrasound image $L$ as an additional network input.
Given the additional ultrasound physics and texture information encoded in the low-quality image input, $LSA2H$ represents an upper-bound (or, considering our architectural improvements, rather a \emph{silver-standard}).

\item[$\bullet$] {\bf NSA2H} a lower-bound baseline by replacing the low-quality image $L$ in $LSA2H$ with an uncorrelated Gaussian noise image $N$.
\end{itemize}

\begin{figure*}
\centering
\ULsubfloat[Target\label{fig:XiParameterCuts0}]{%
       \includegraphics[width=.22\textwidth]{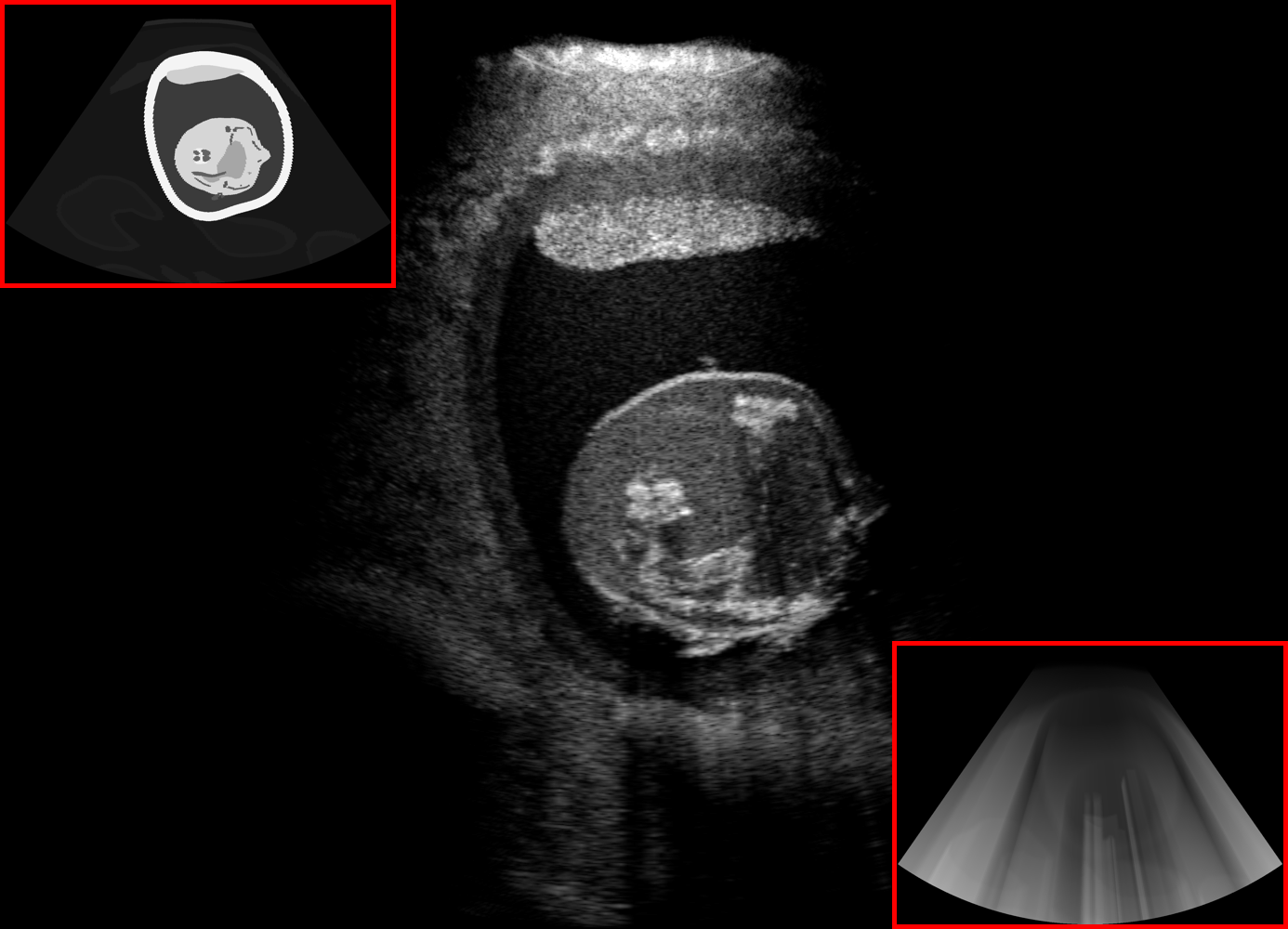}
\vspace{10pt}
\includegraphics[width=.22\textwidth]{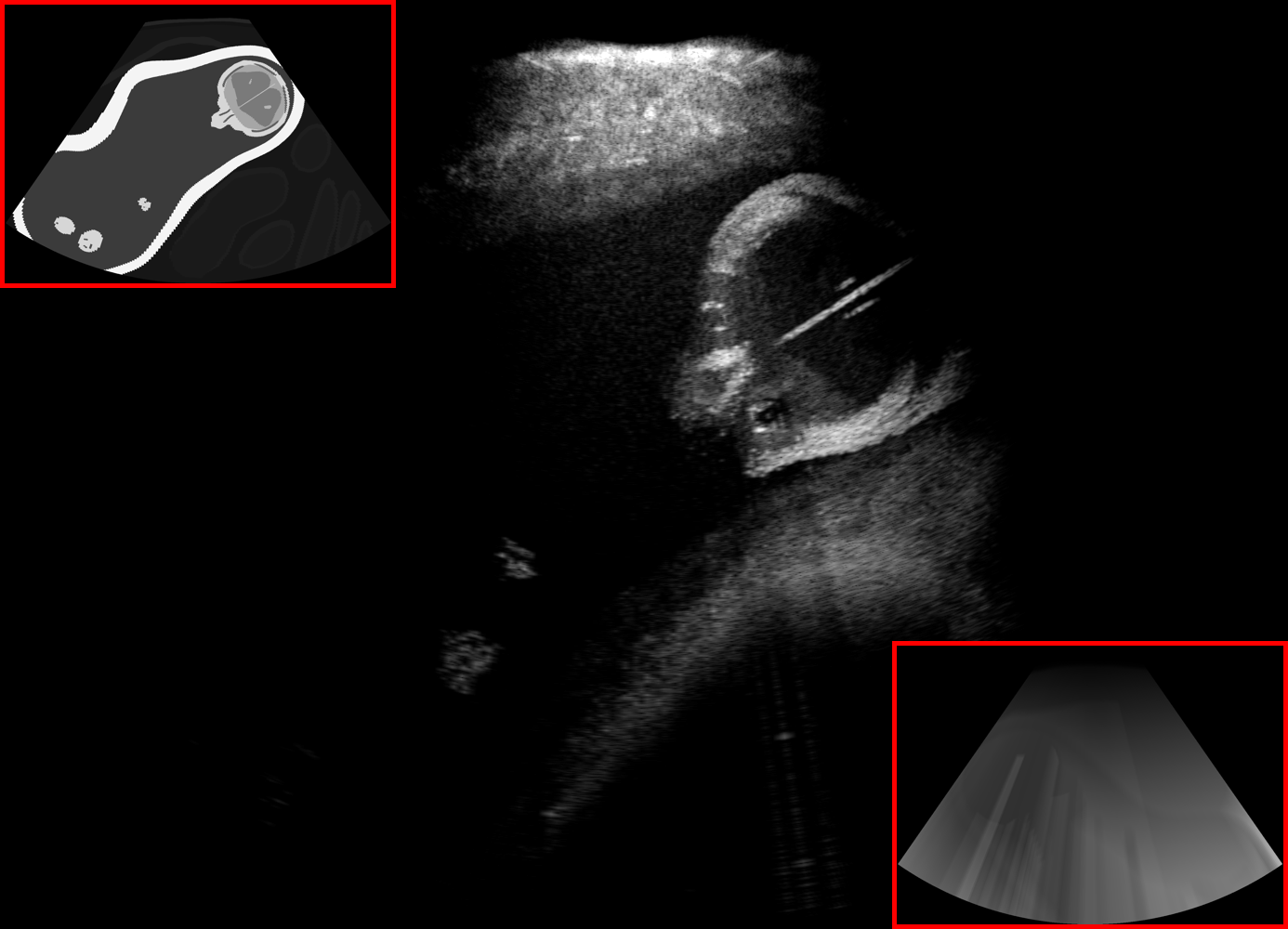}
\includegraphics[width=.22\textwidth]{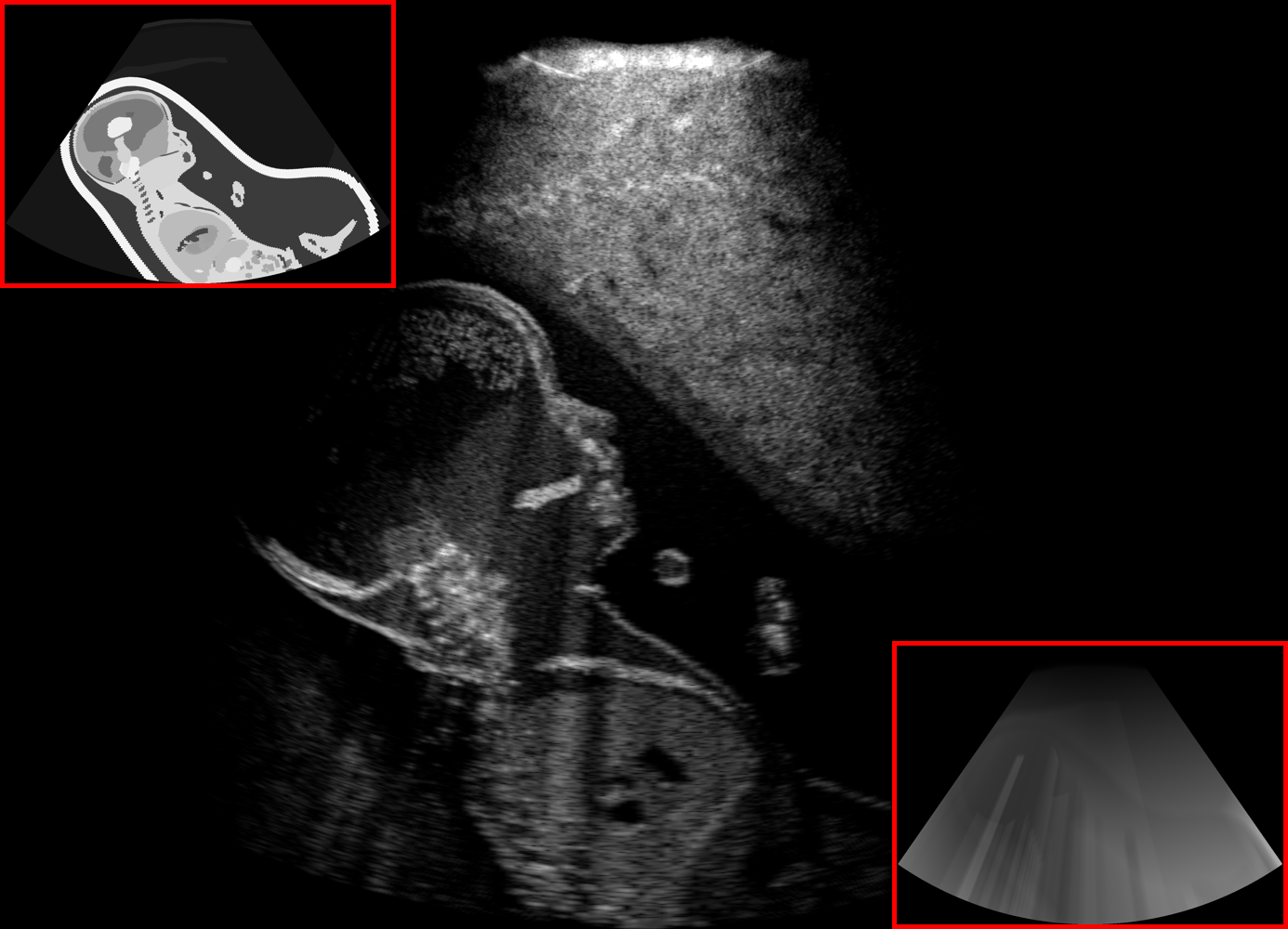}
\includegraphics[width=.22\textwidth]{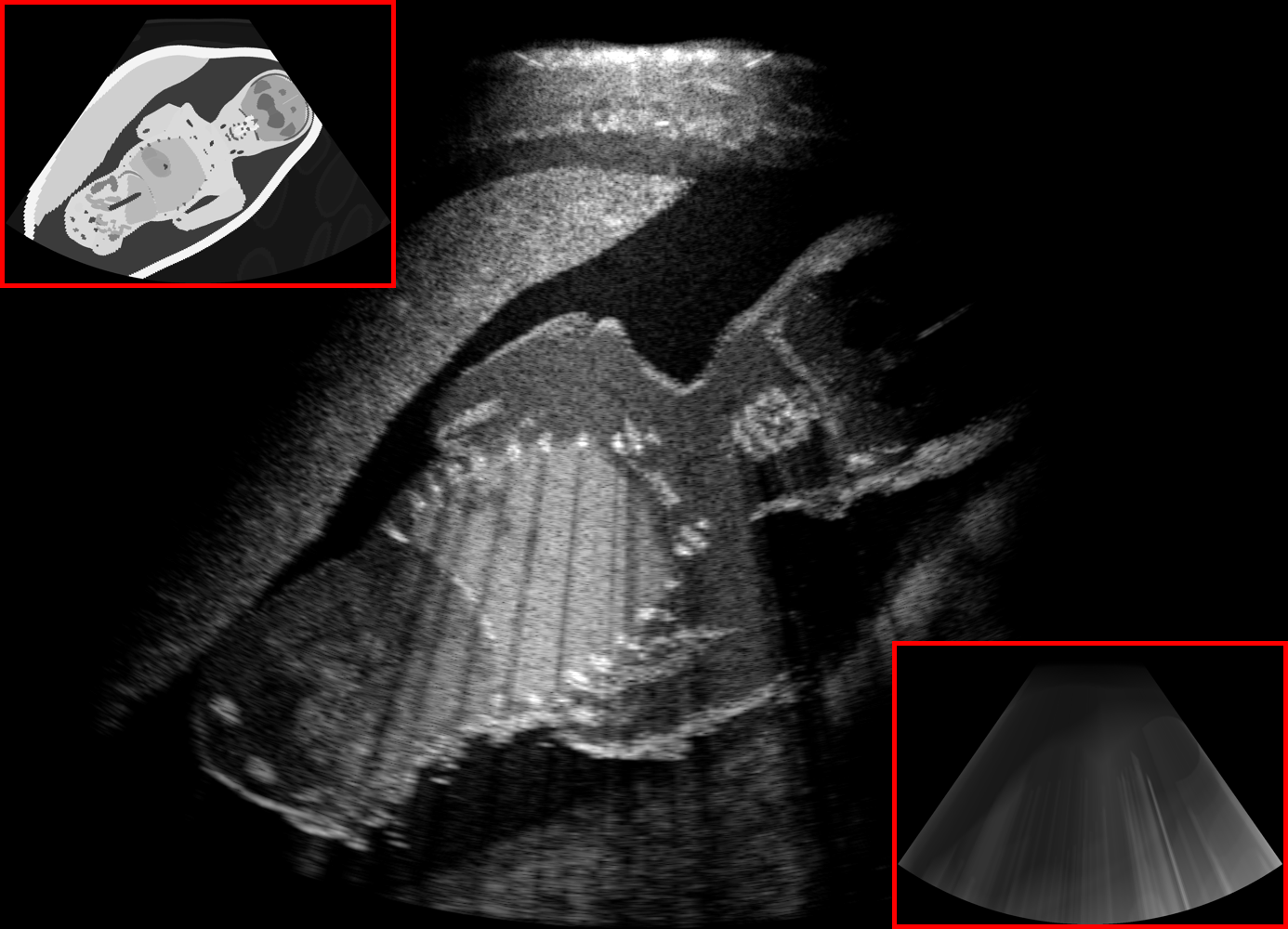}}

\centering
\ULsubfloat[SA2H\label{fig:XiParameterCuts1}]{%
       \includegraphics[width=.22\textwidth]{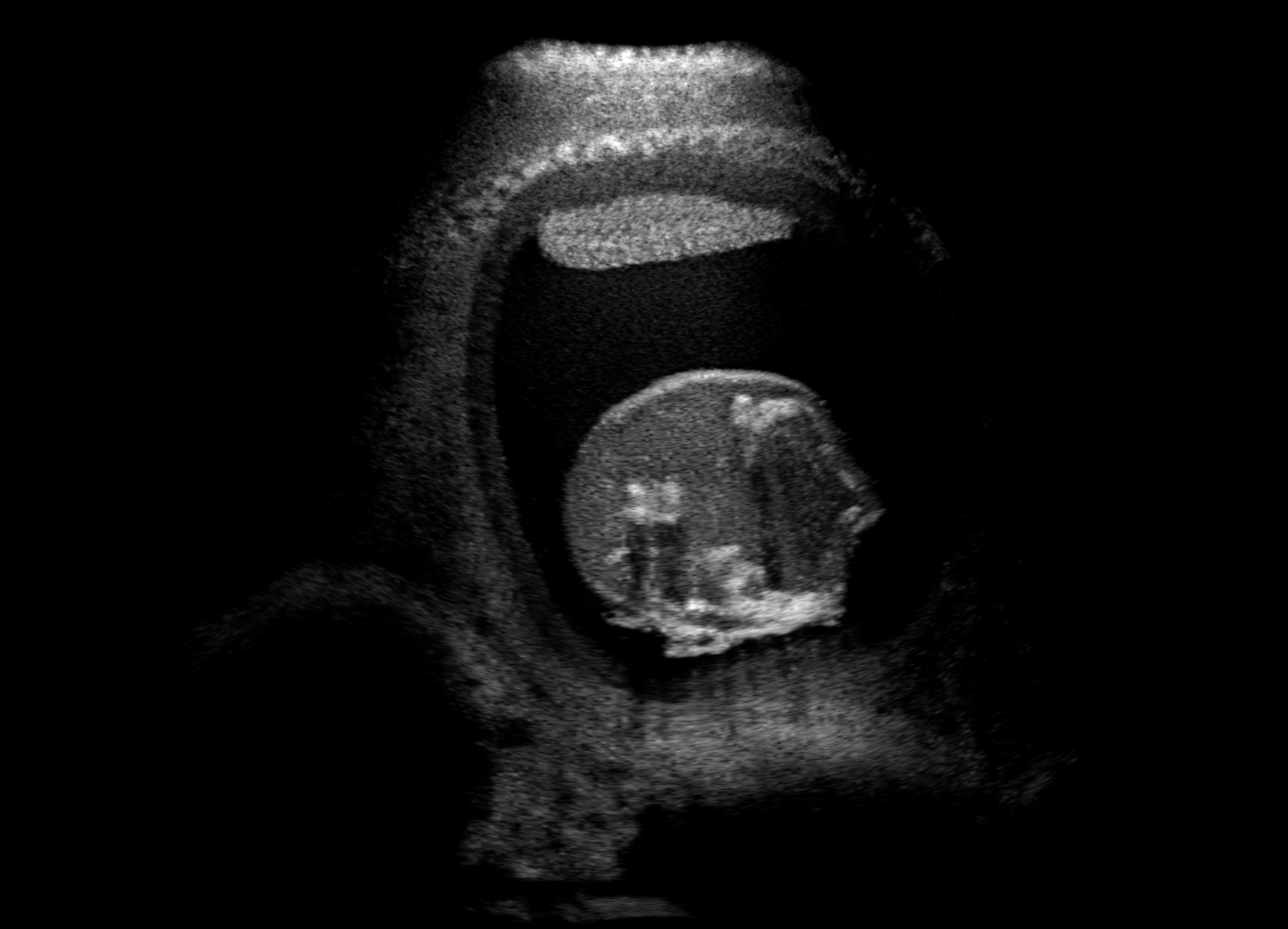}
\vspace{10pt}
\includegraphics[width=.22\textwidth]{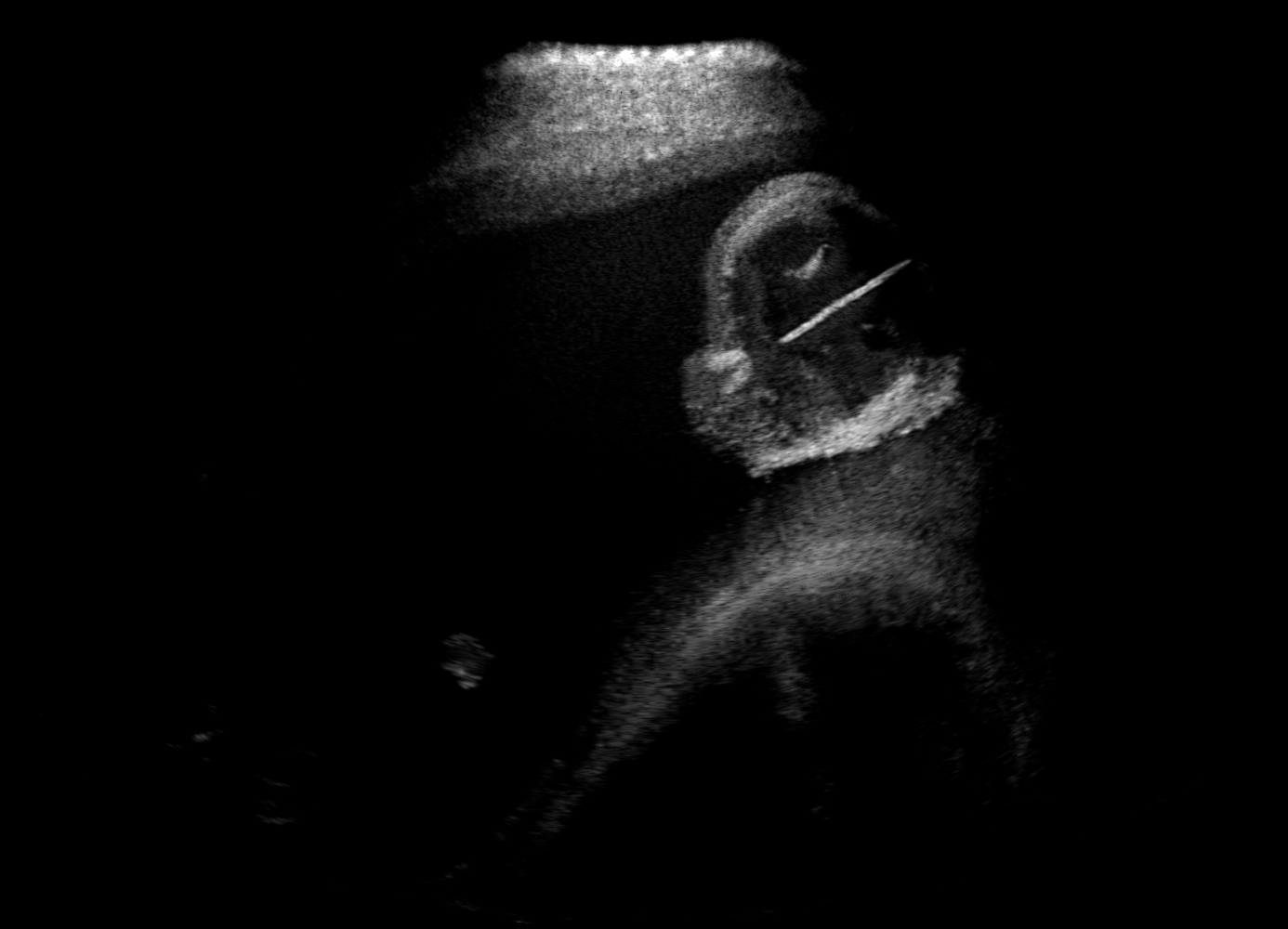}
\includegraphics[width=.22\textwidth]{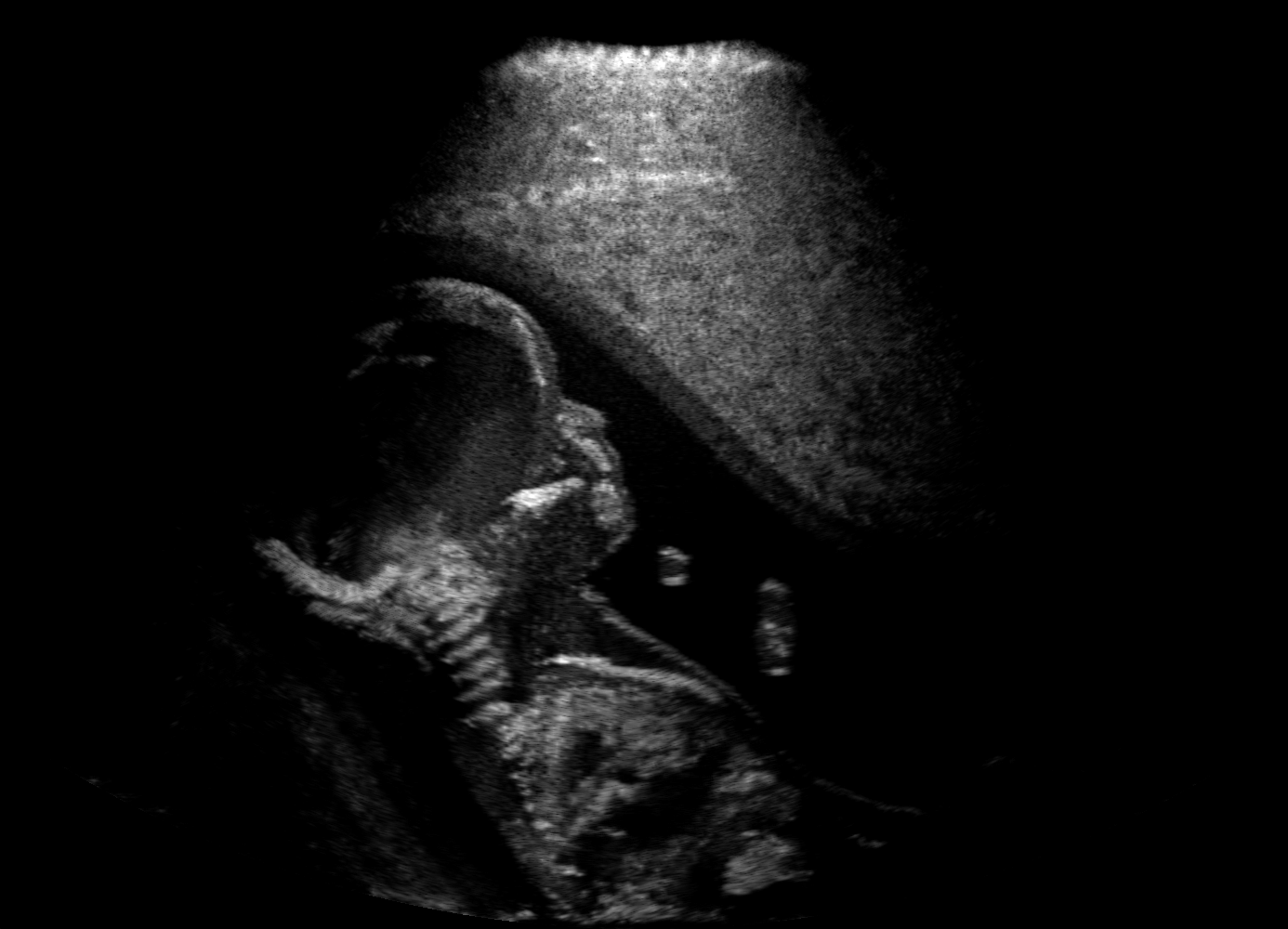}
\includegraphics[width=.22\textwidth]{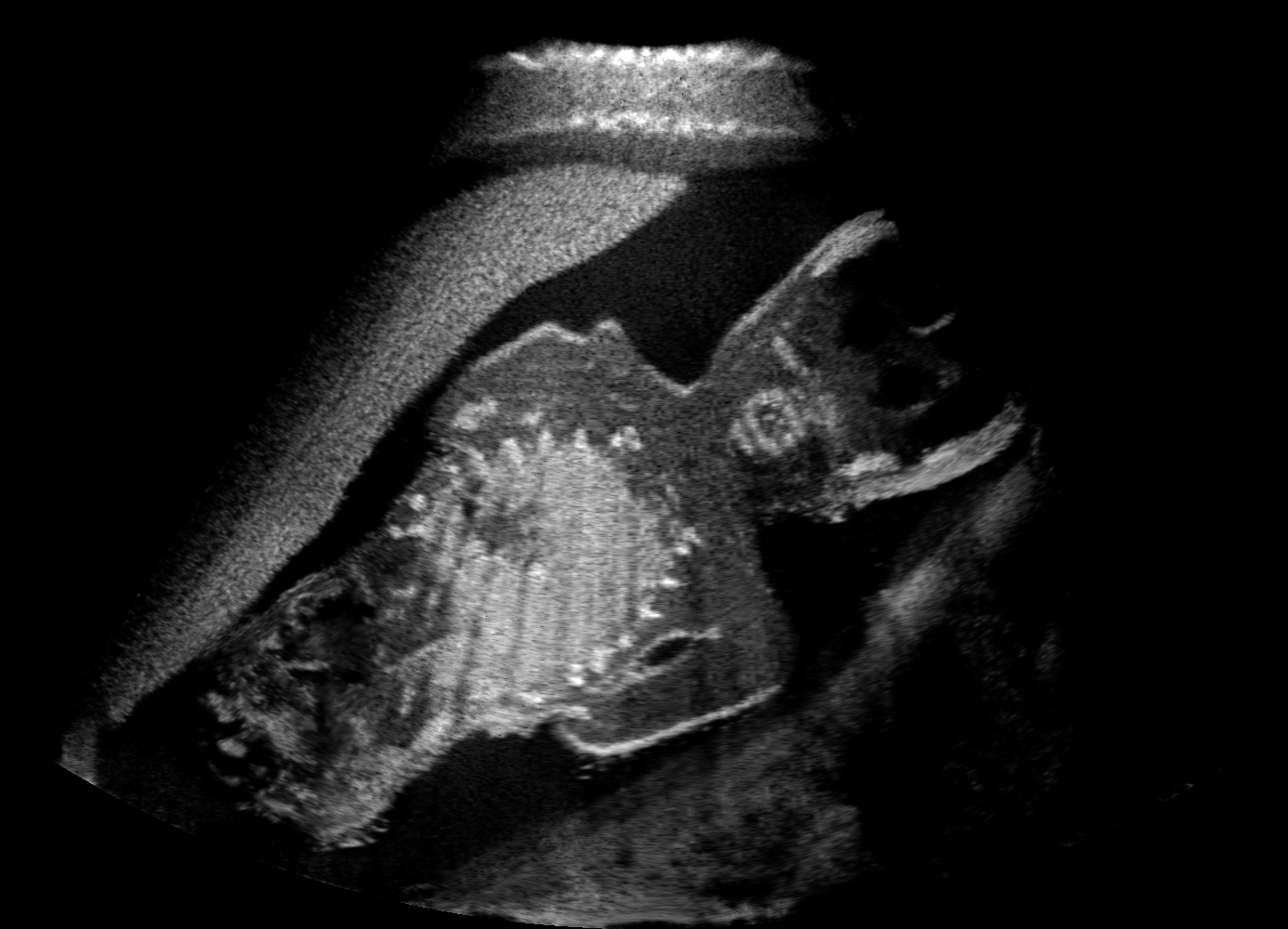}}

\centering
\ULsubfloat[SA2H-att\label{fig:XiParameterCuts2}]{%
       \includegraphics[width=.22\textwidth]{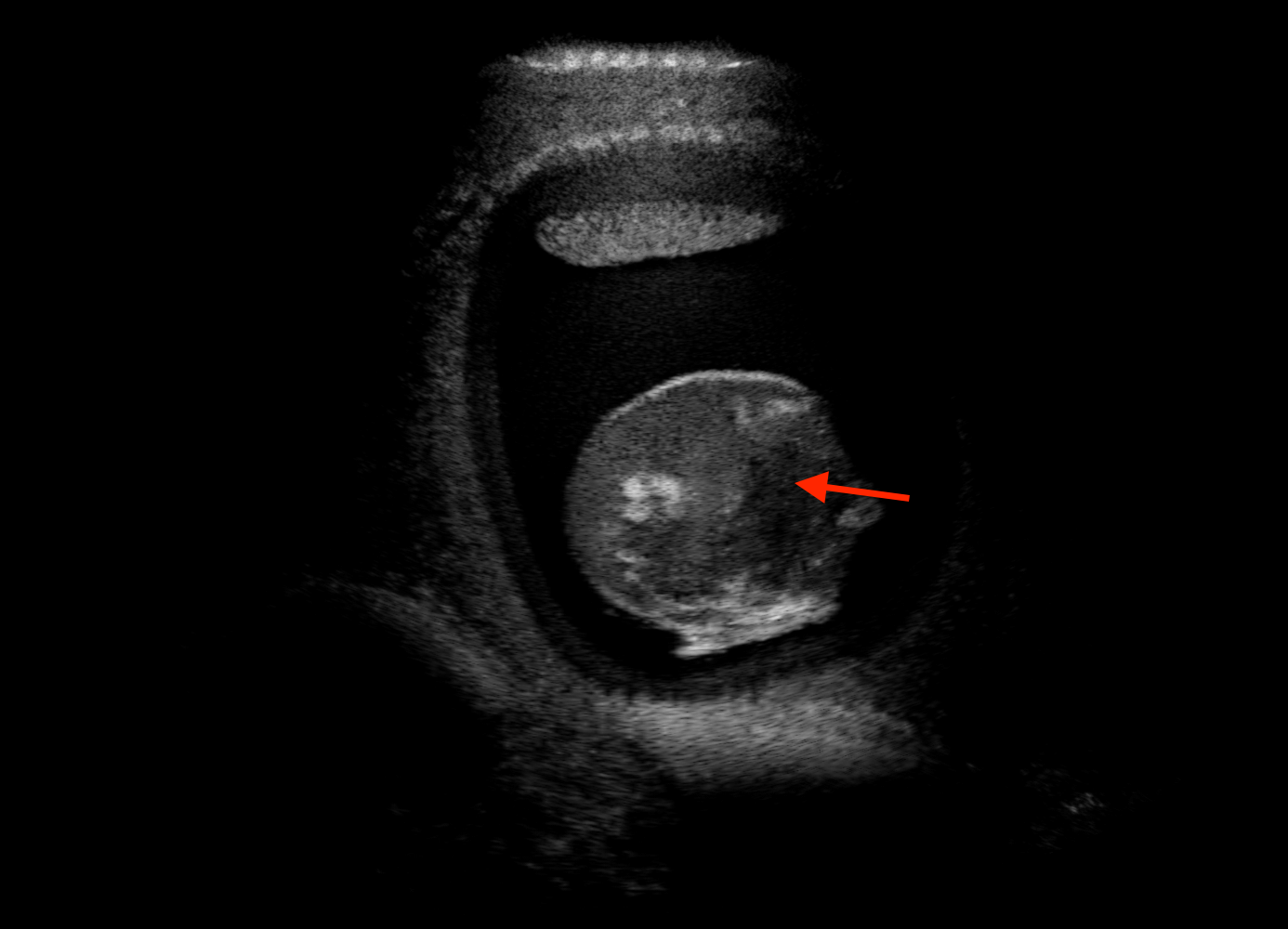}
\vspace{10pt}
\includegraphics[width=.22\textwidth]{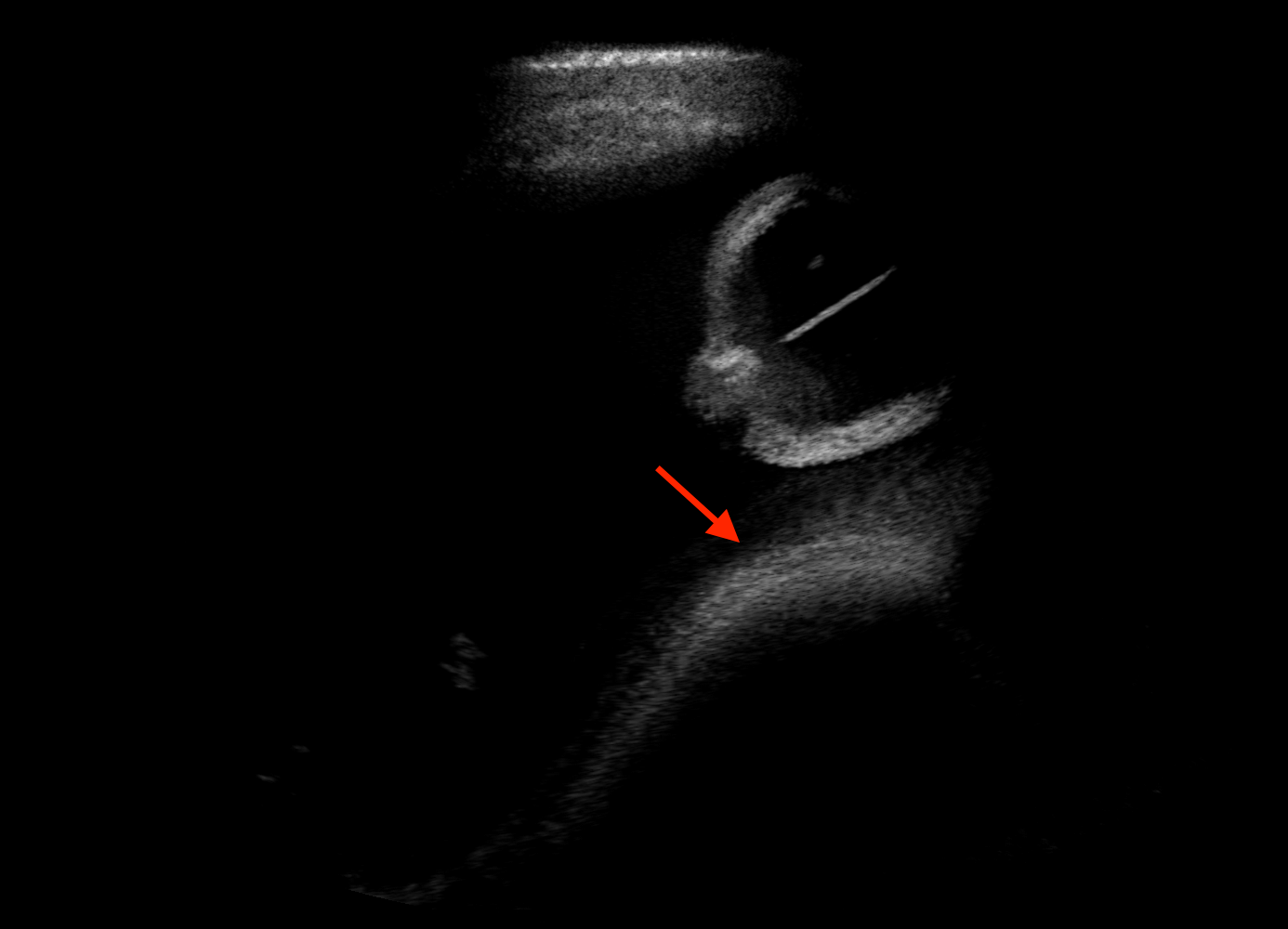}
\includegraphics[width=.22\textwidth]{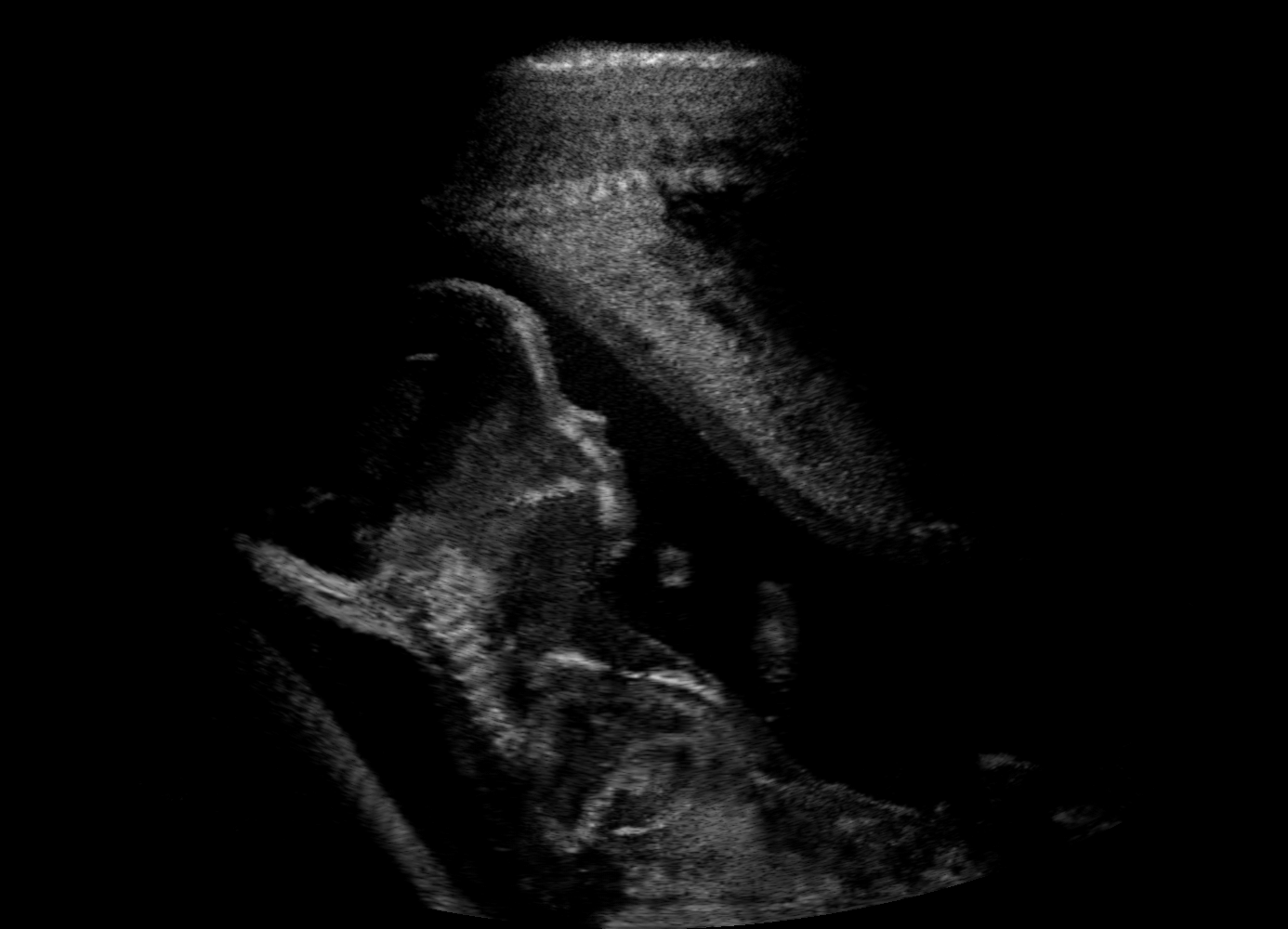}
\includegraphics[width=.22\textwidth]{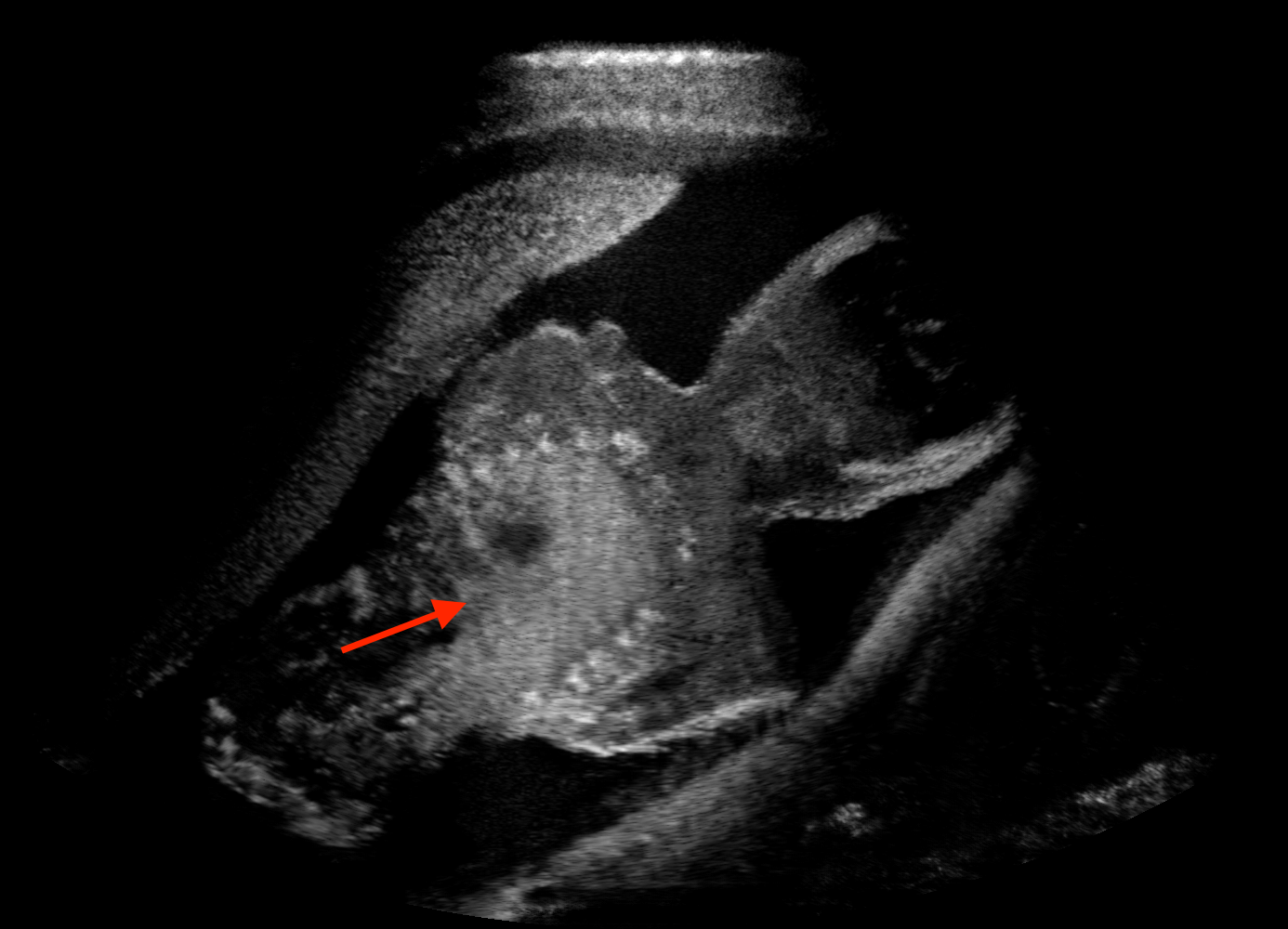}}

\centering
\ULsubfloat[SA2H-concat\label{fig:XiParameterCuts3}]{%
       \includegraphics[width=.22\textwidth]{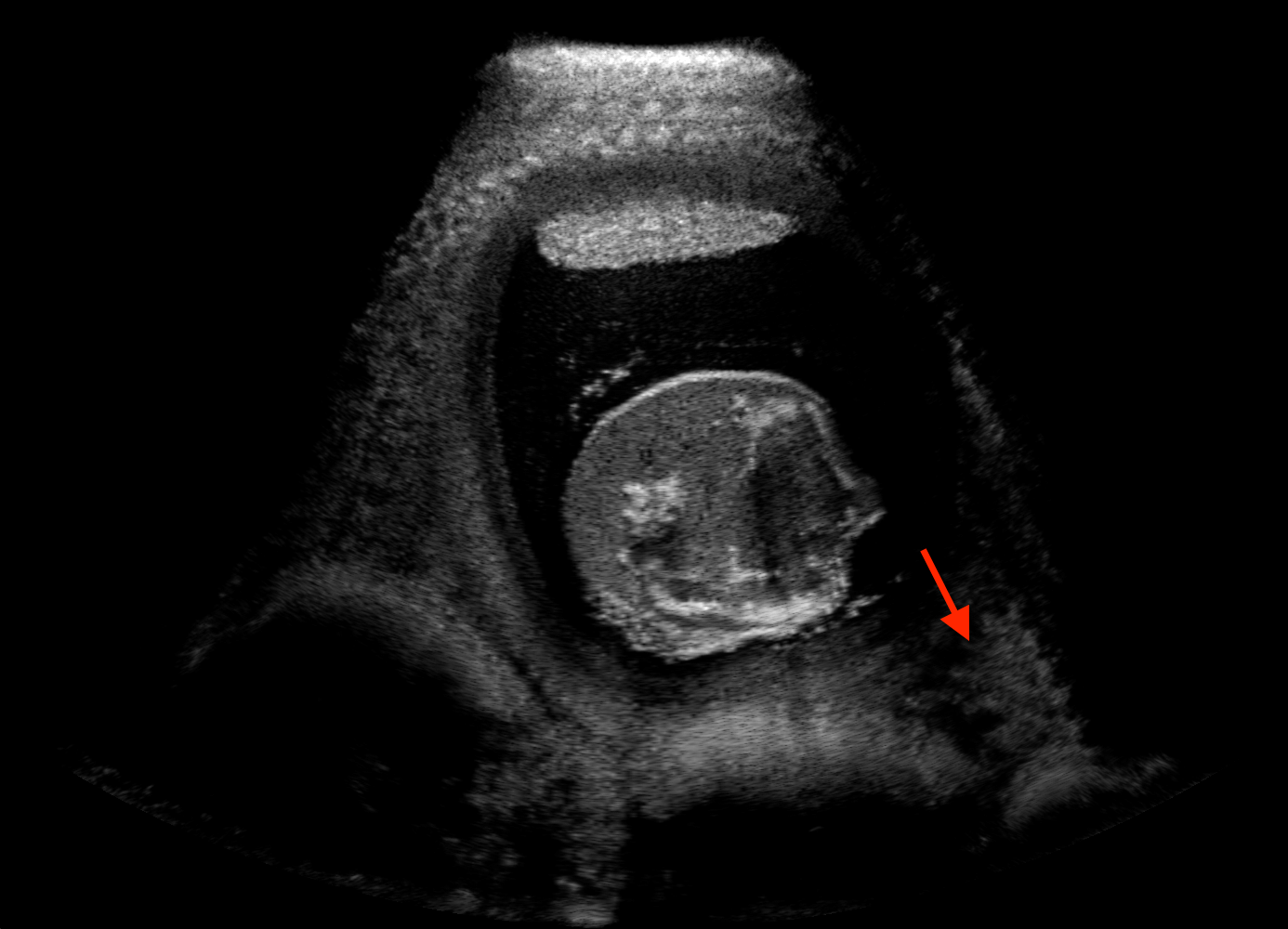}
\vspace{10pt}
\includegraphics[width=.22\textwidth]{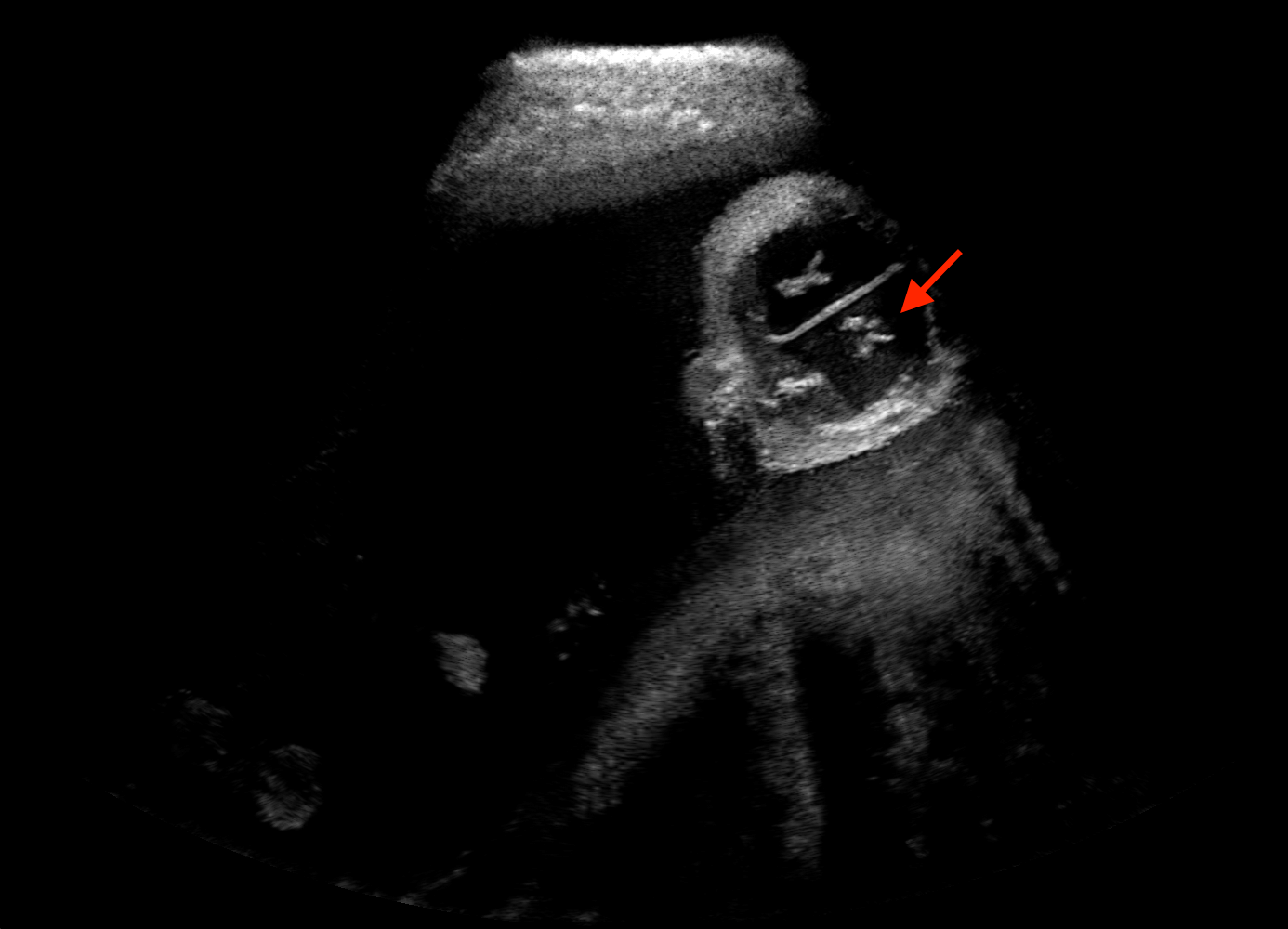}
\includegraphics[width=.22\textwidth]{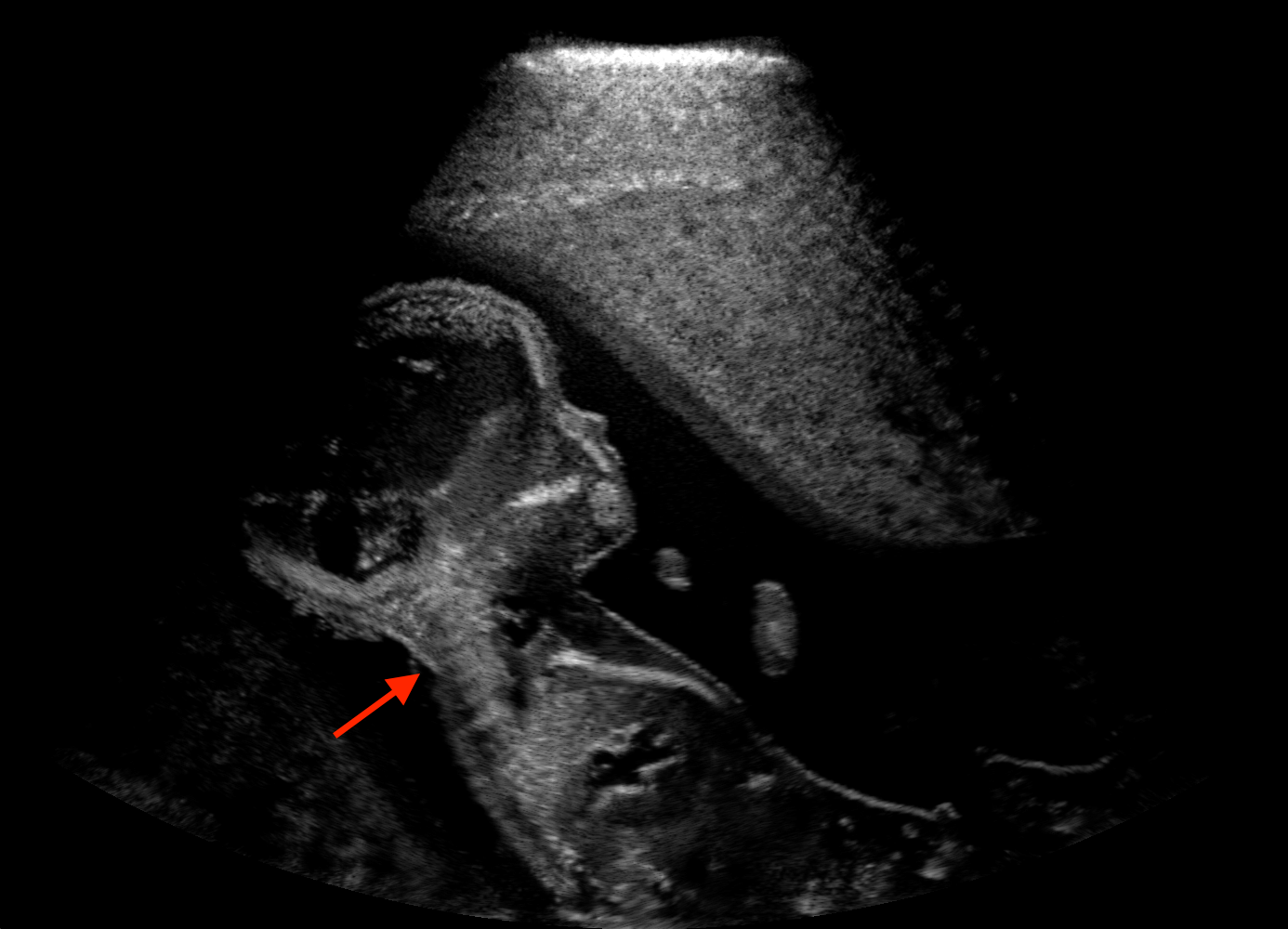}
\includegraphics[width=.22\textwidth]{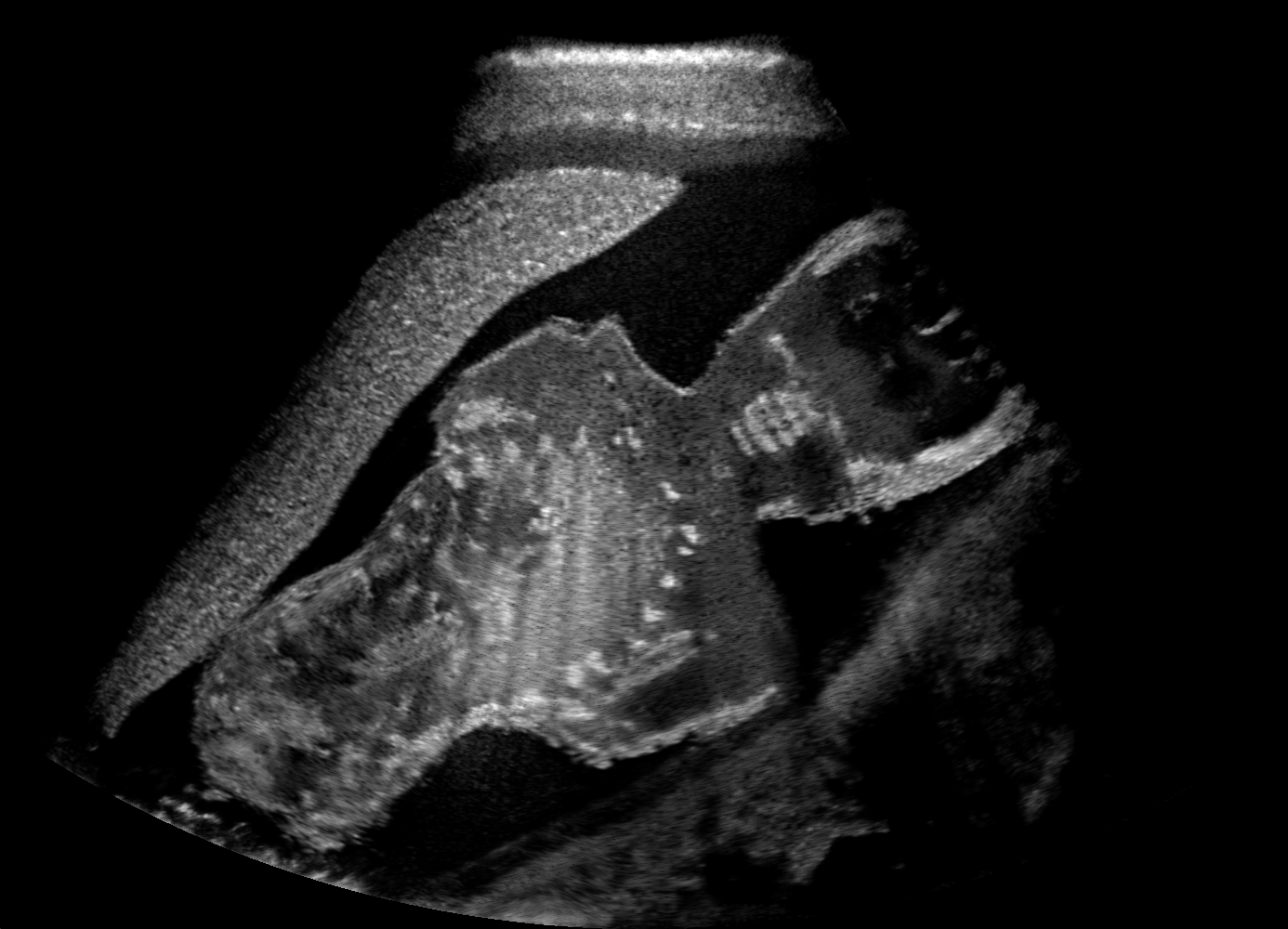}}

\centering
\ULsubfloat[SA2H-conv\label{fig:XiParameterCuts4}]{%
       \includegraphics[width=.22\textwidth]{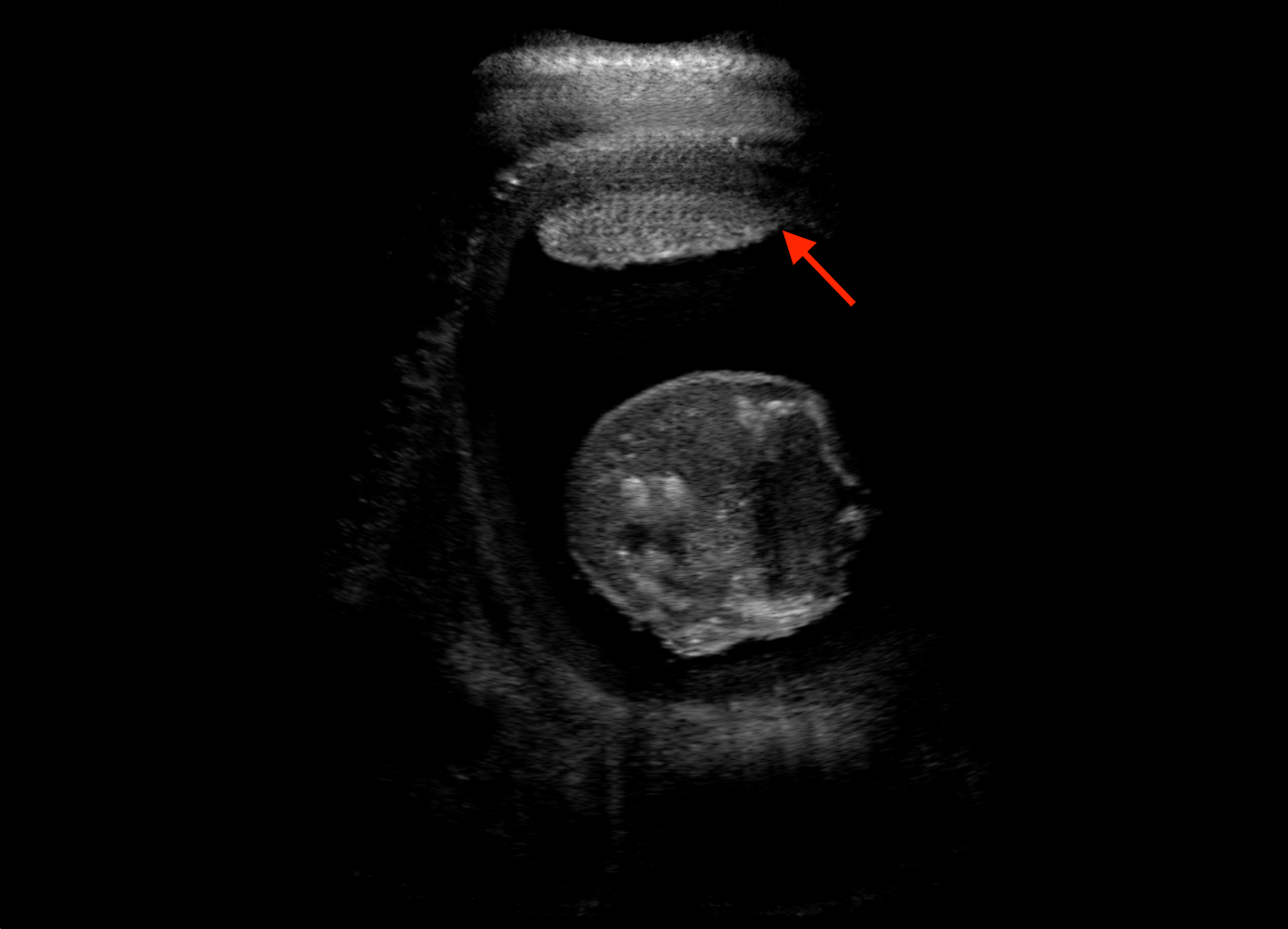}
\vspace{10pt}
\includegraphics[width=.22\textwidth]{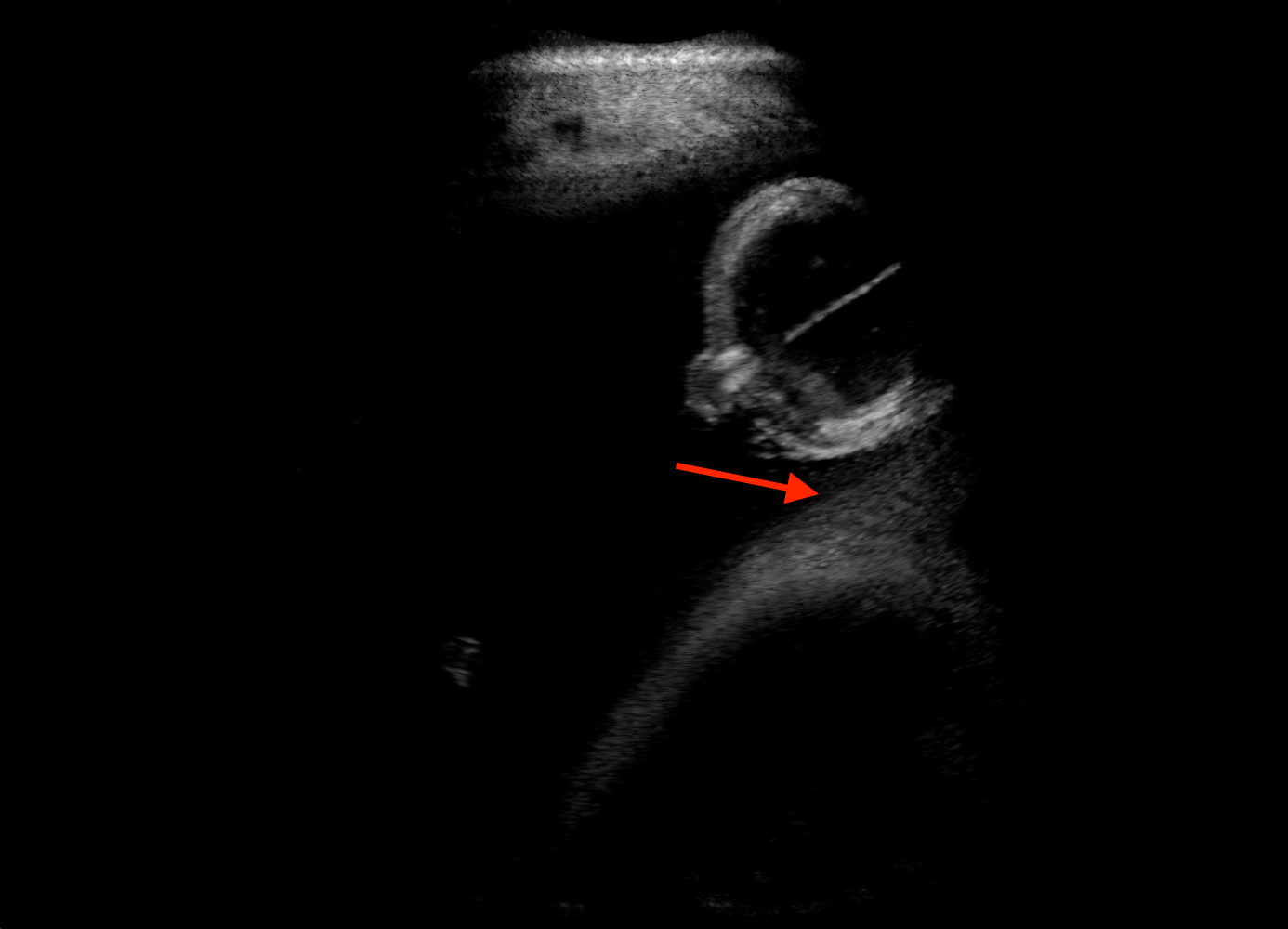}
\includegraphics[width=.22\textwidth]{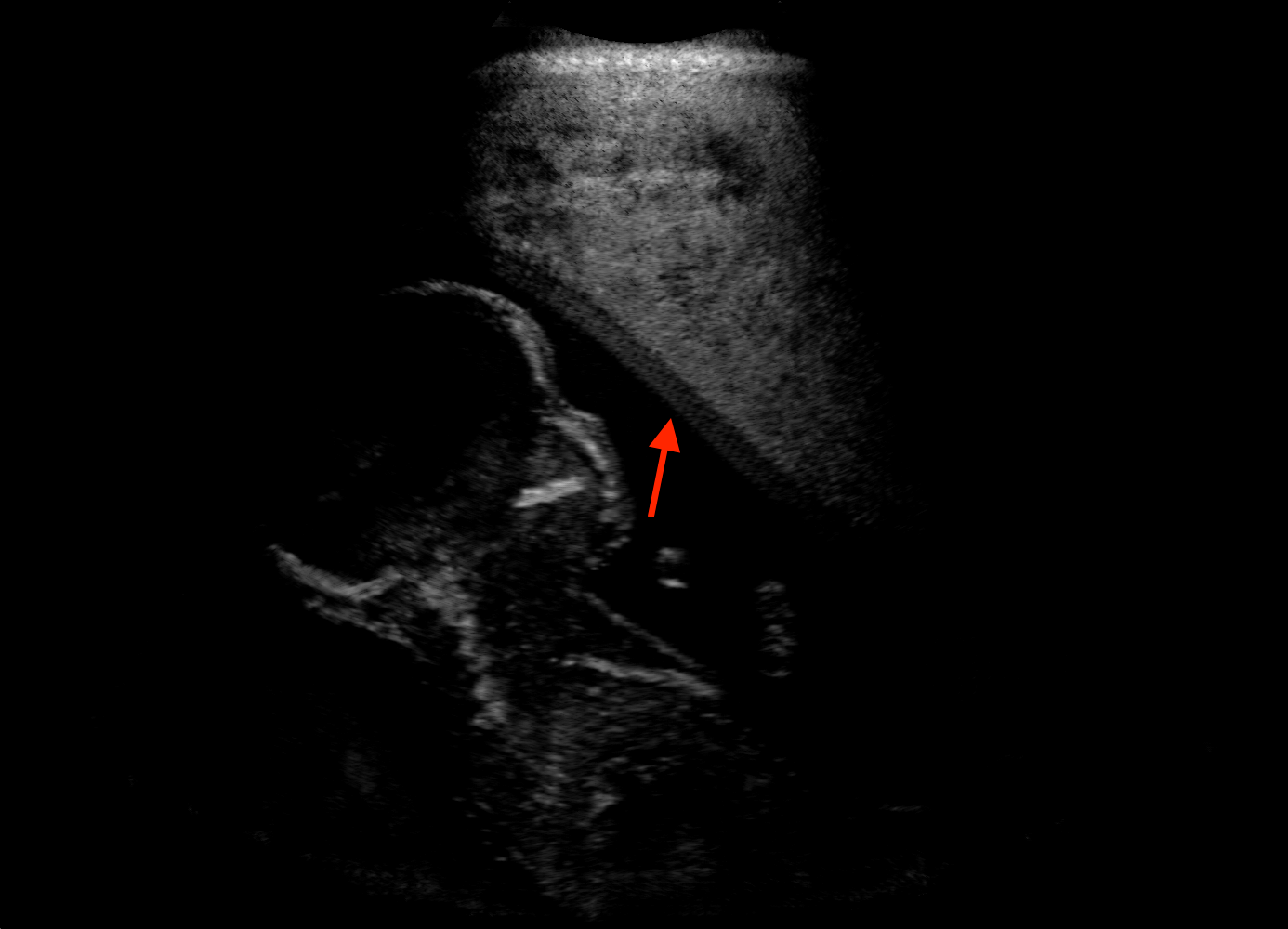}
\includegraphics[width=.22\textwidth]{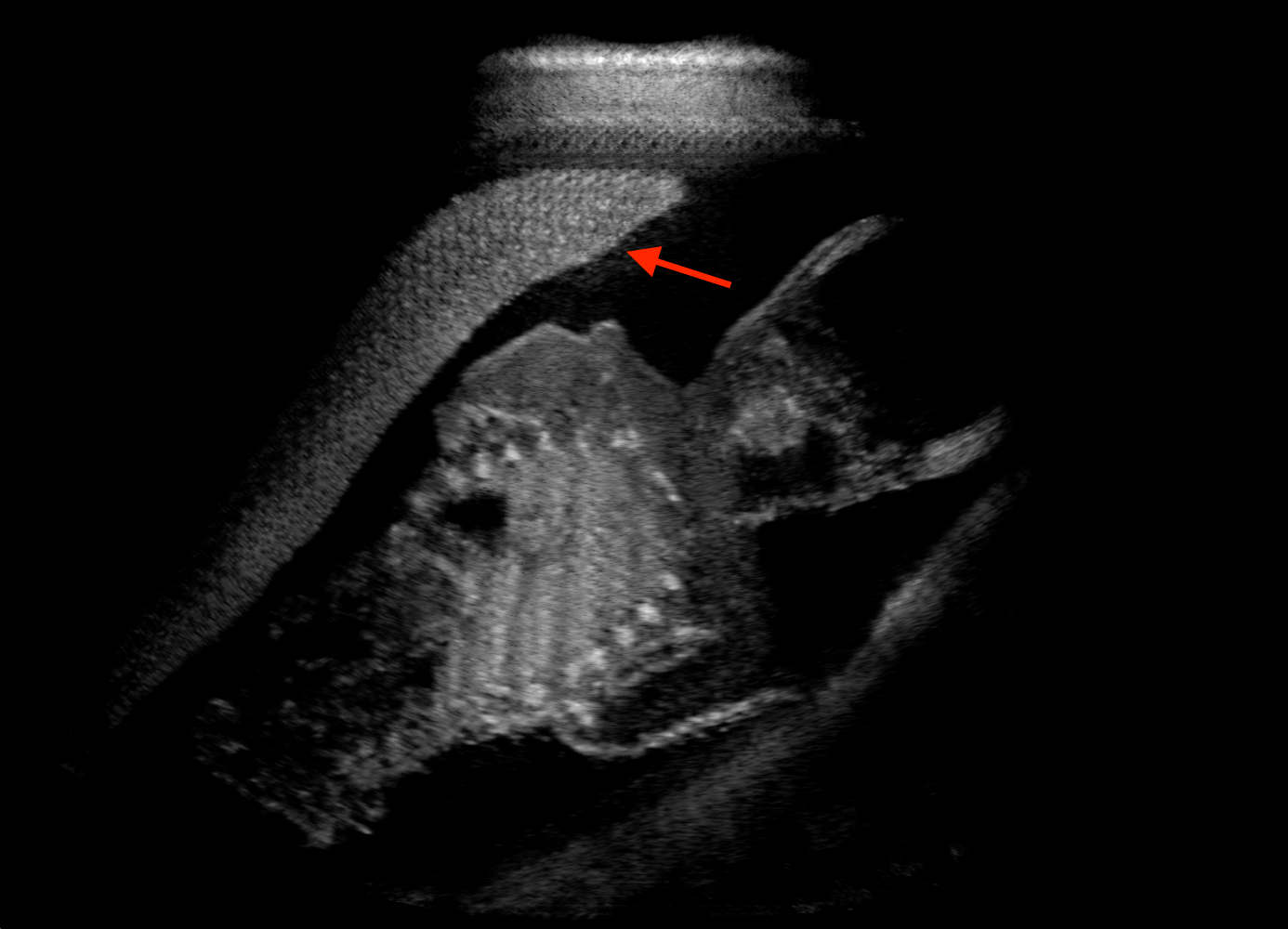}}

\centering
\ULsubfloat[SA2H-noise\label{fig:XiParameterCuts5}]{%
       \includegraphics[width=.22\textwidth]{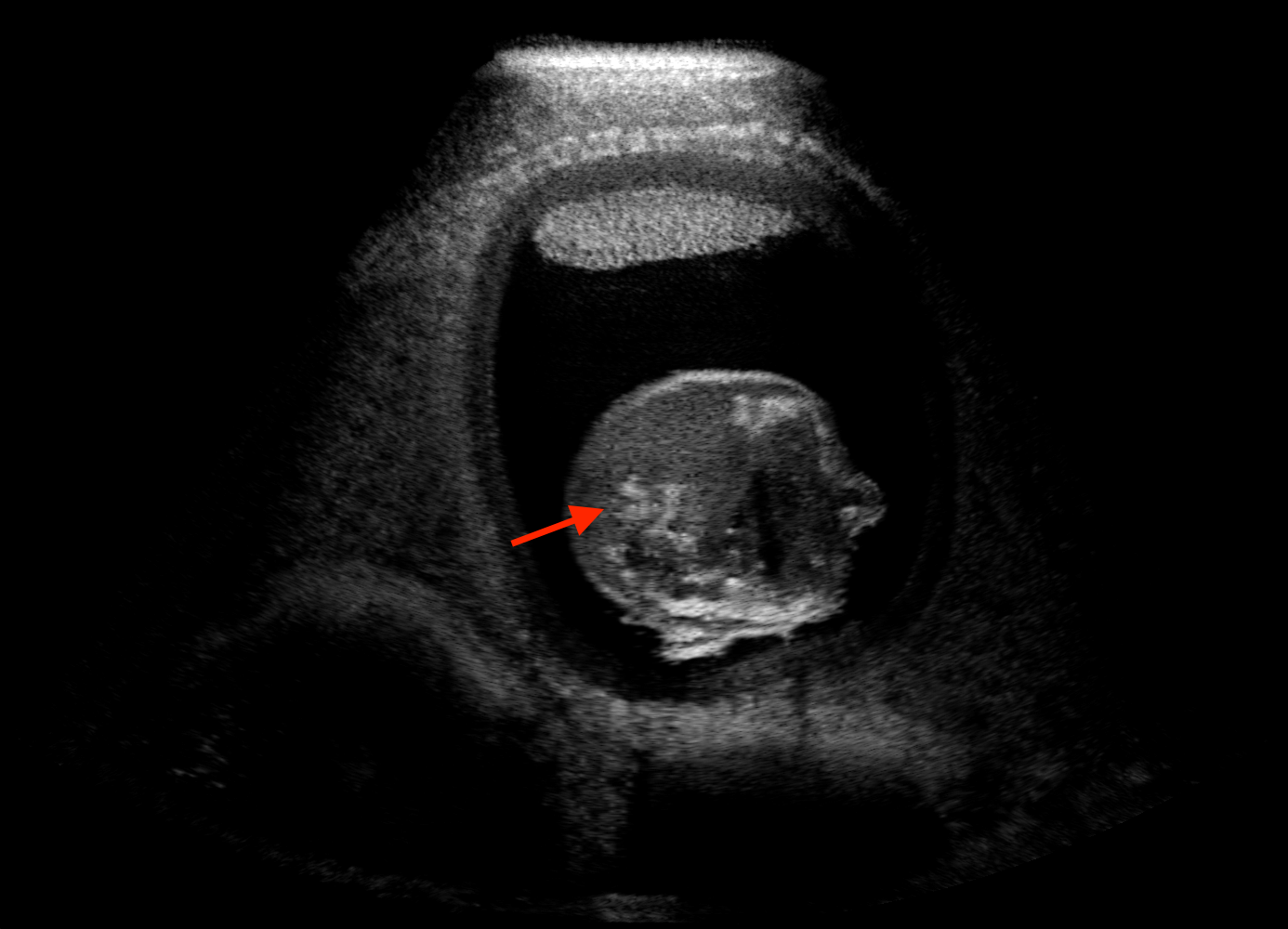}
\vspace{10pt}
\includegraphics[width=.22\textwidth]{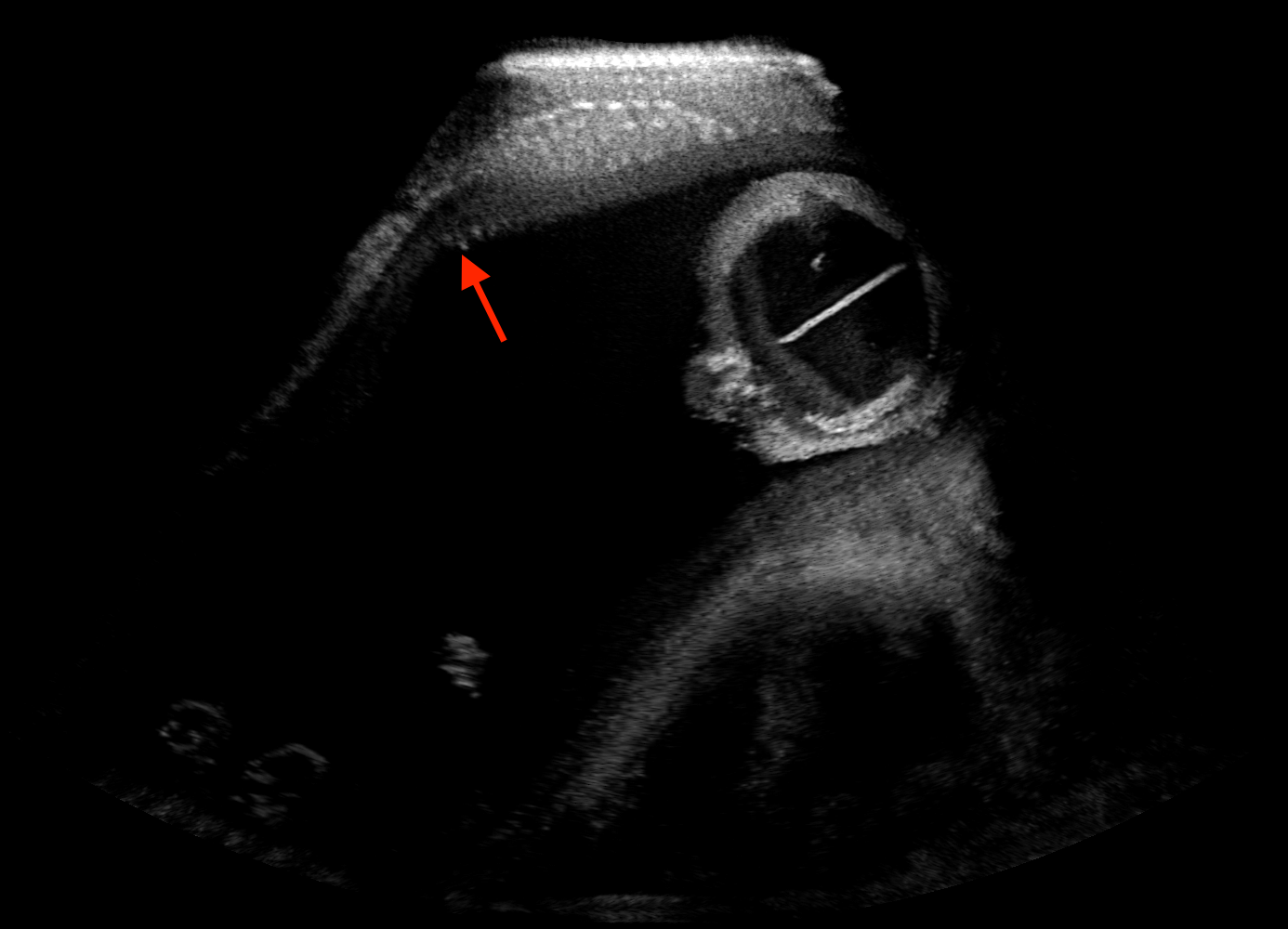}
\includegraphics[width=.22\textwidth]{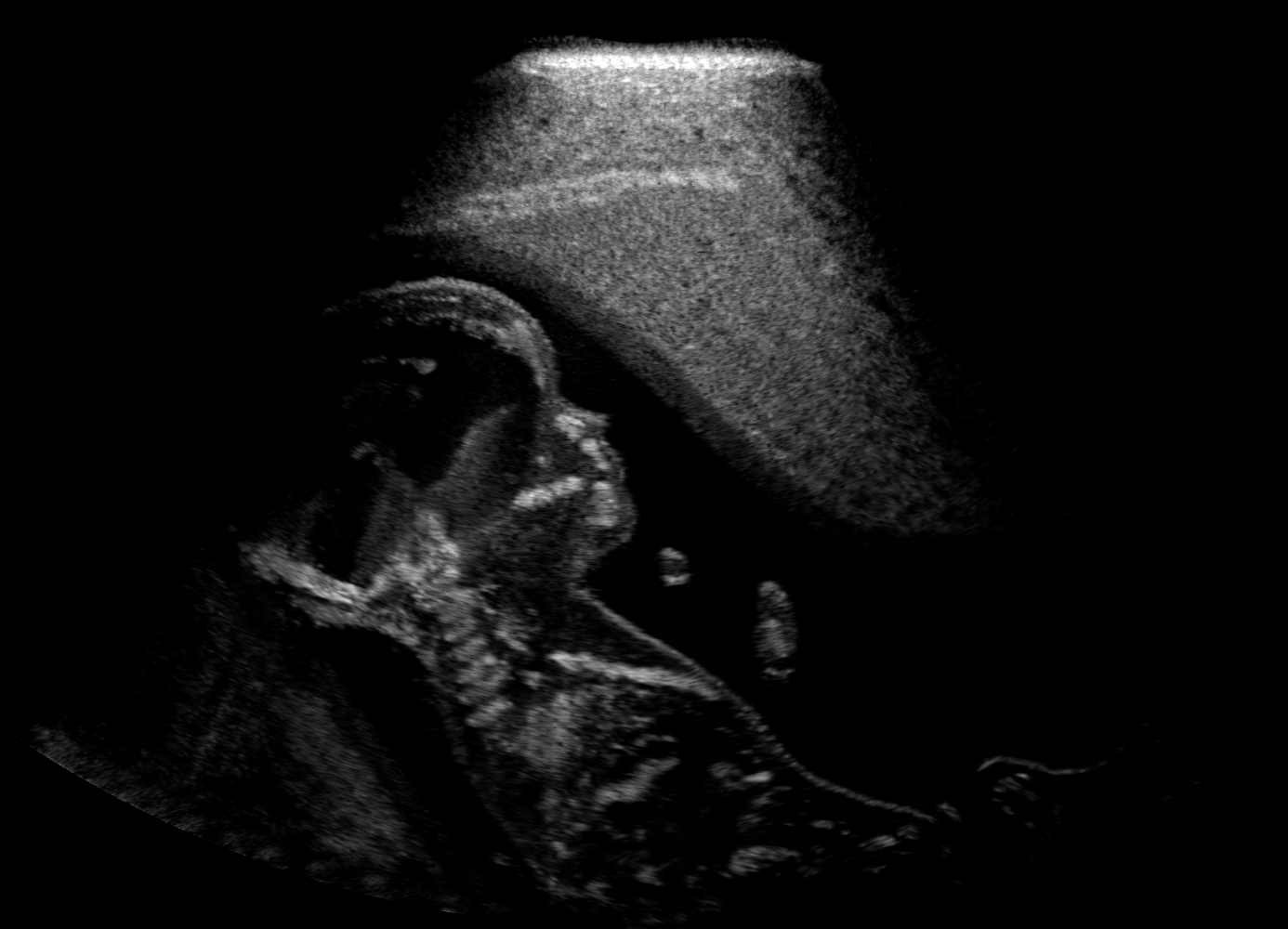}
\includegraphics[width=.22\textwidth]{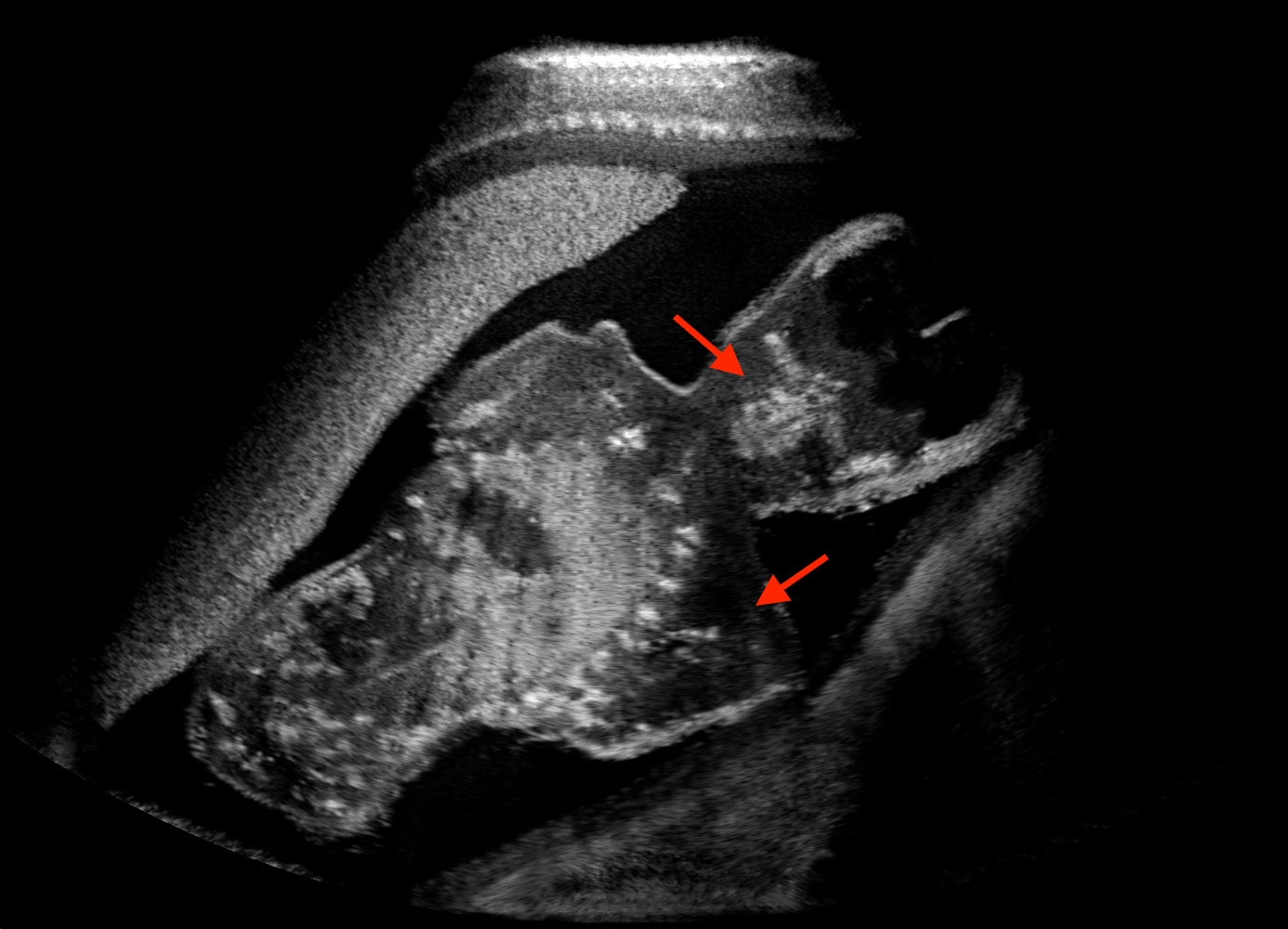}}

\centering
\ULsubfloat[NSA2H\label{fig:XiParameterCuts6}]{%
       \includegraphics[width=.22\textwidth]{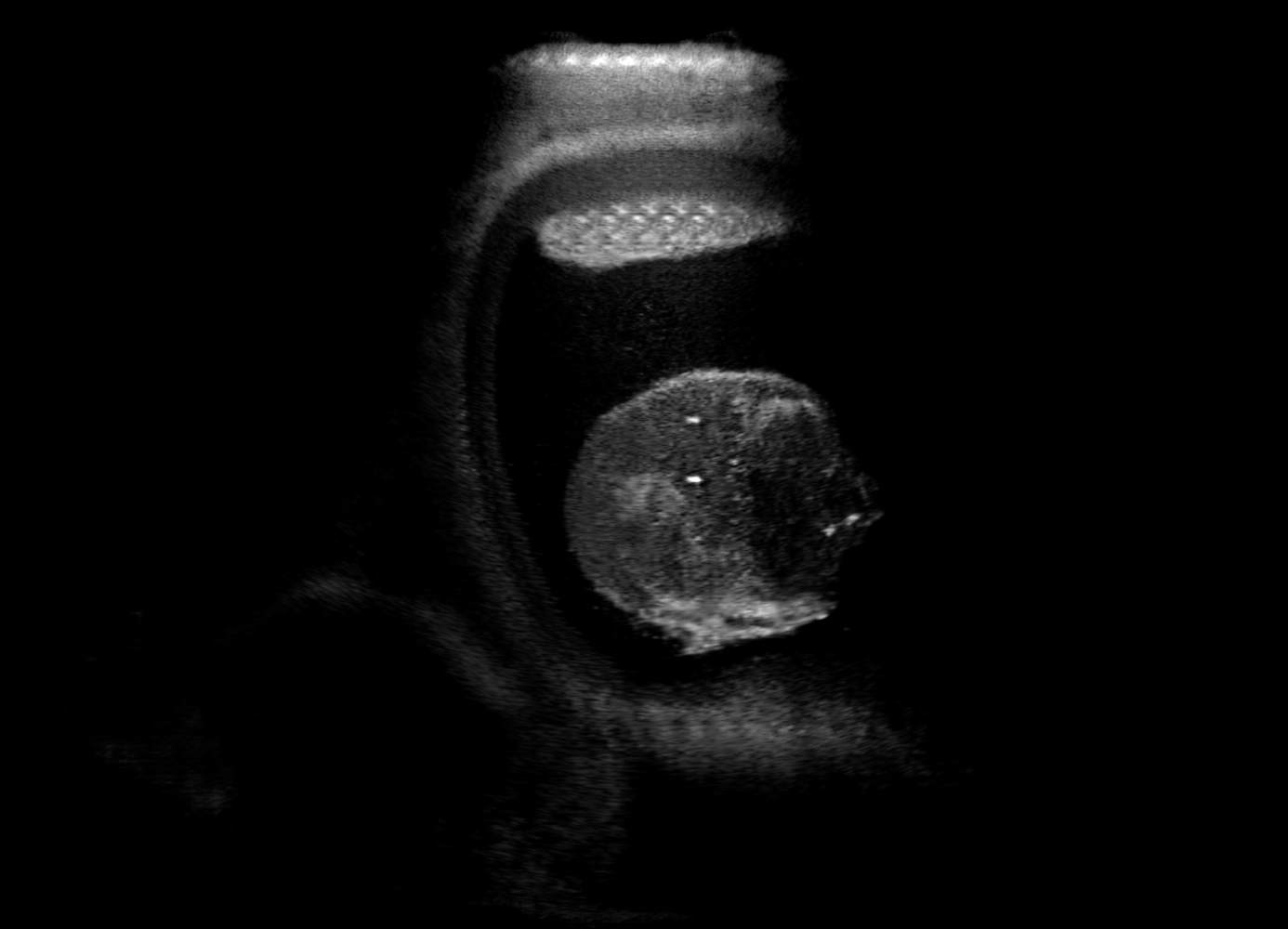}
\vspace{10pt}
\includegraphics[width=.22\textwidth]{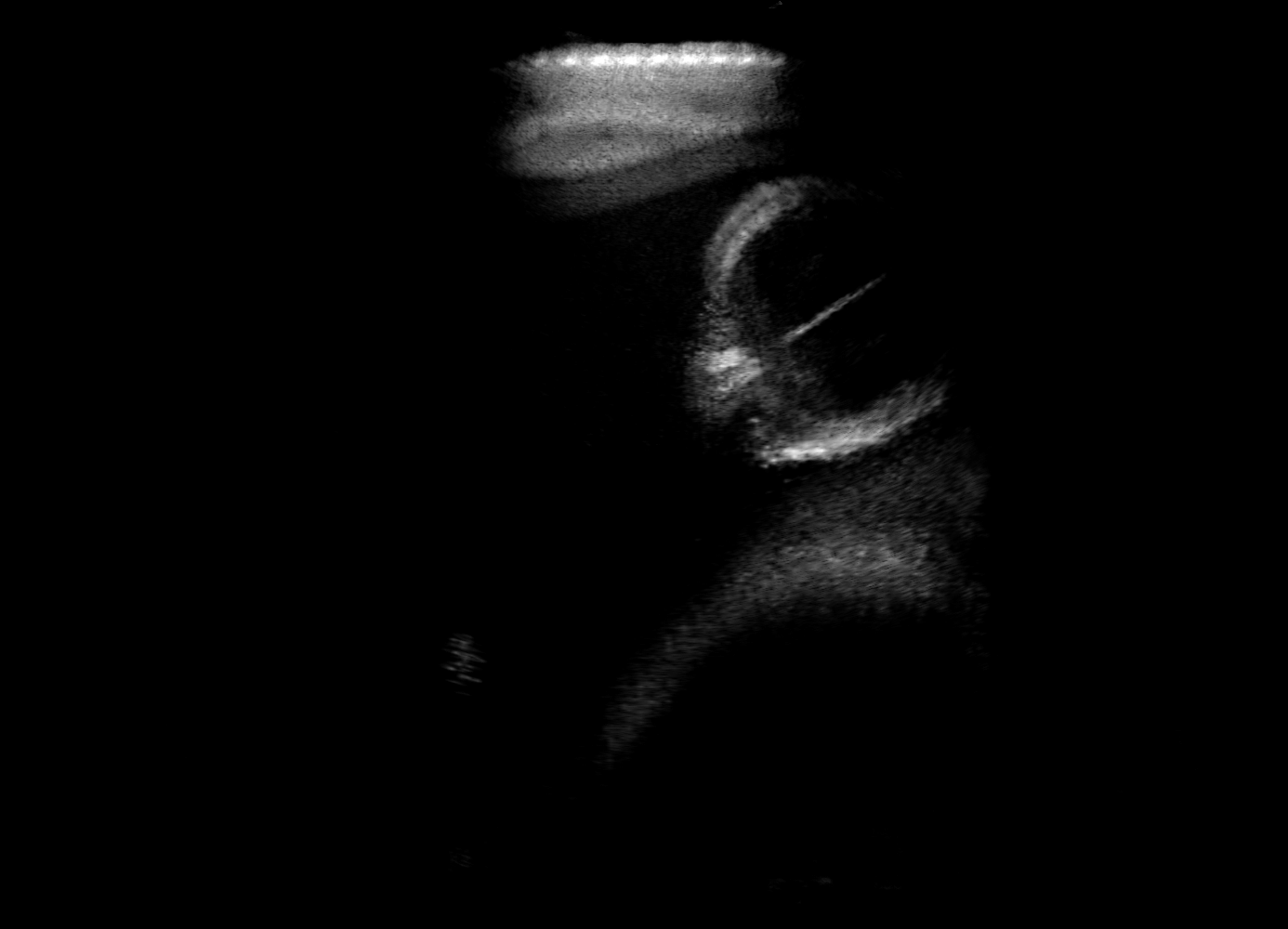}
\includegraphics[width=.22\textwidth]{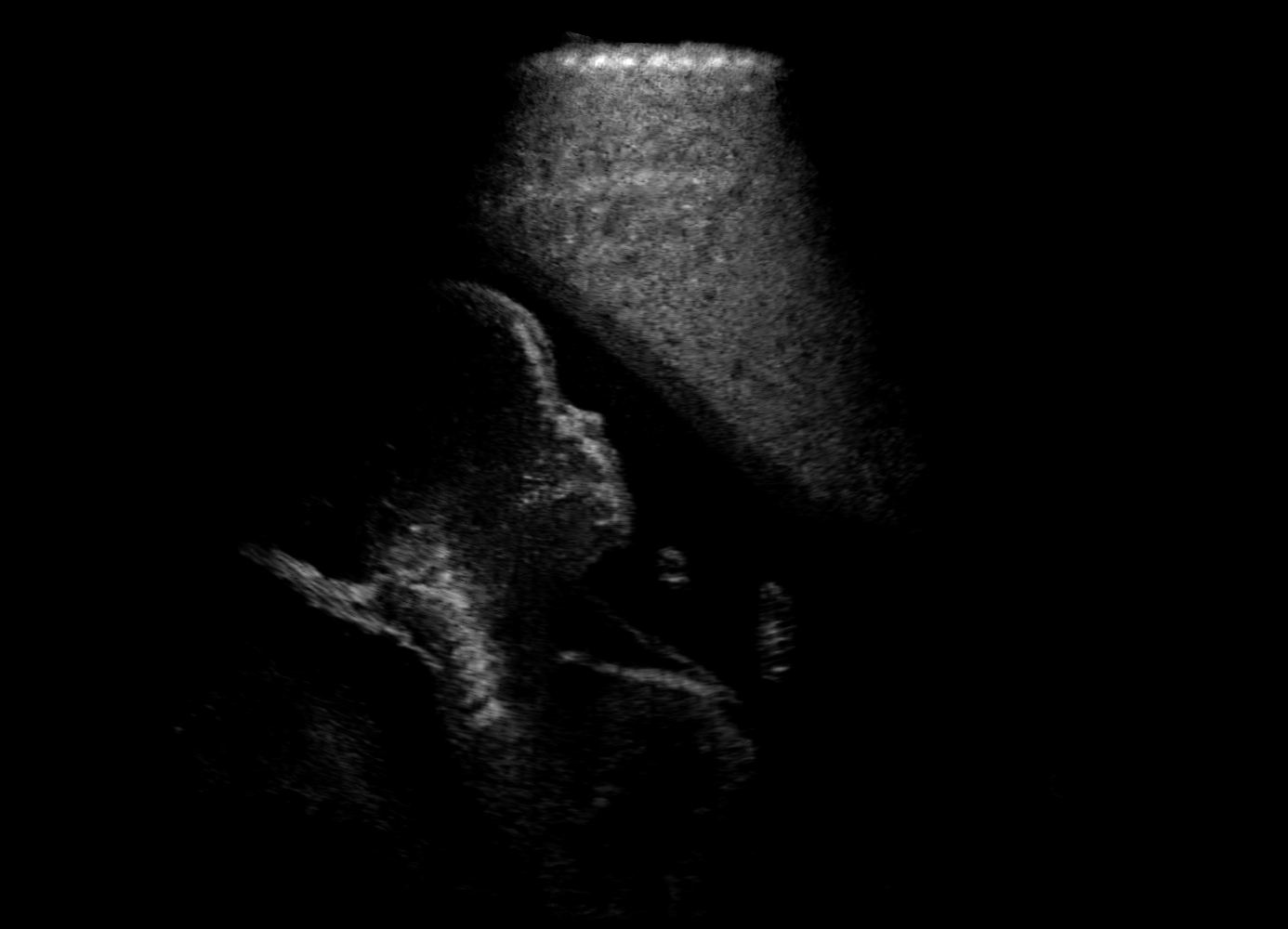}
\includegraphics[width=.22\textwidth]{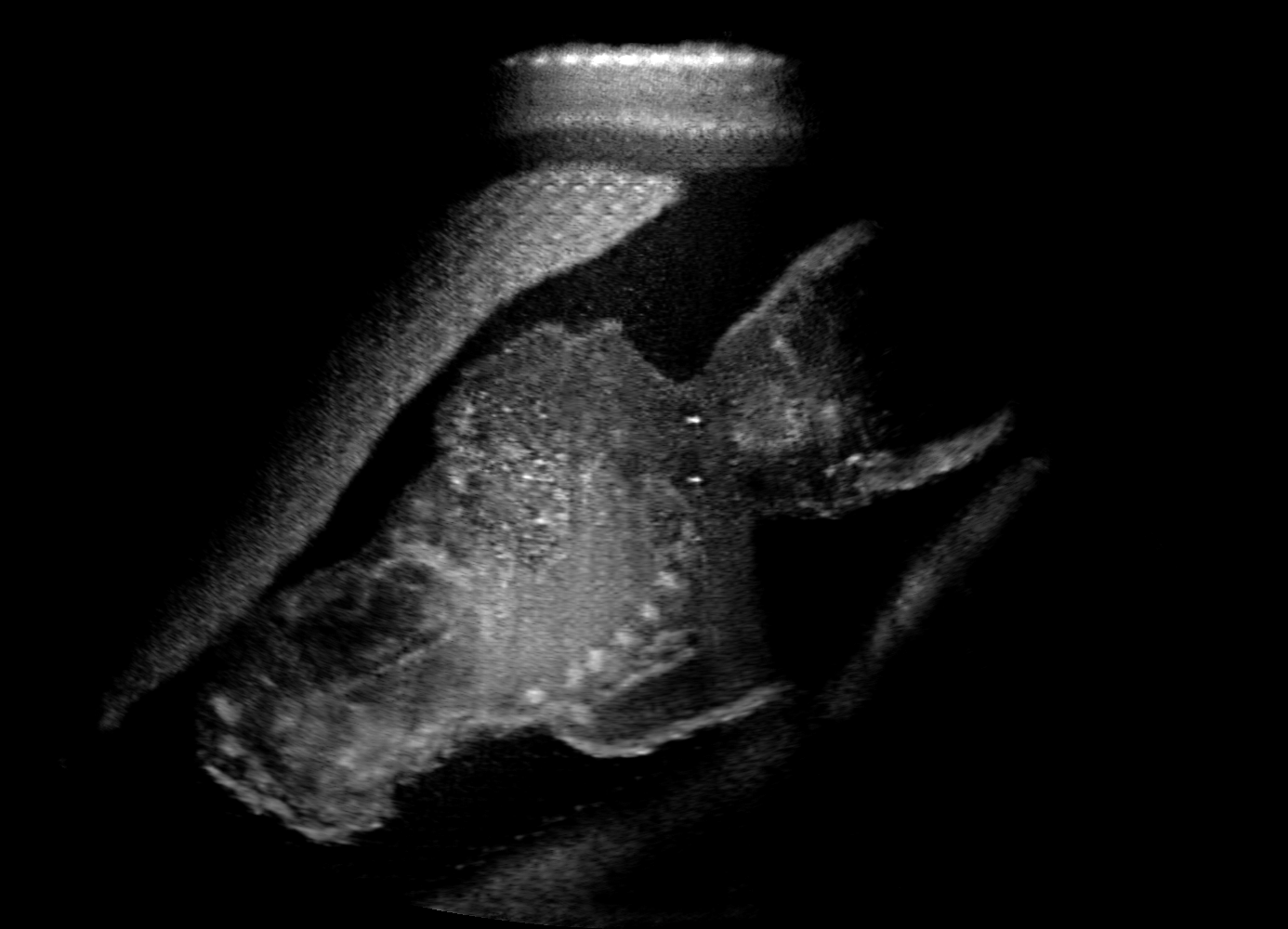}}

\centering
\ULsubfloat[LSA2H\label{fig:XiParameterCuts7}]{%
       \includegraphics[width=.22\textwidth]{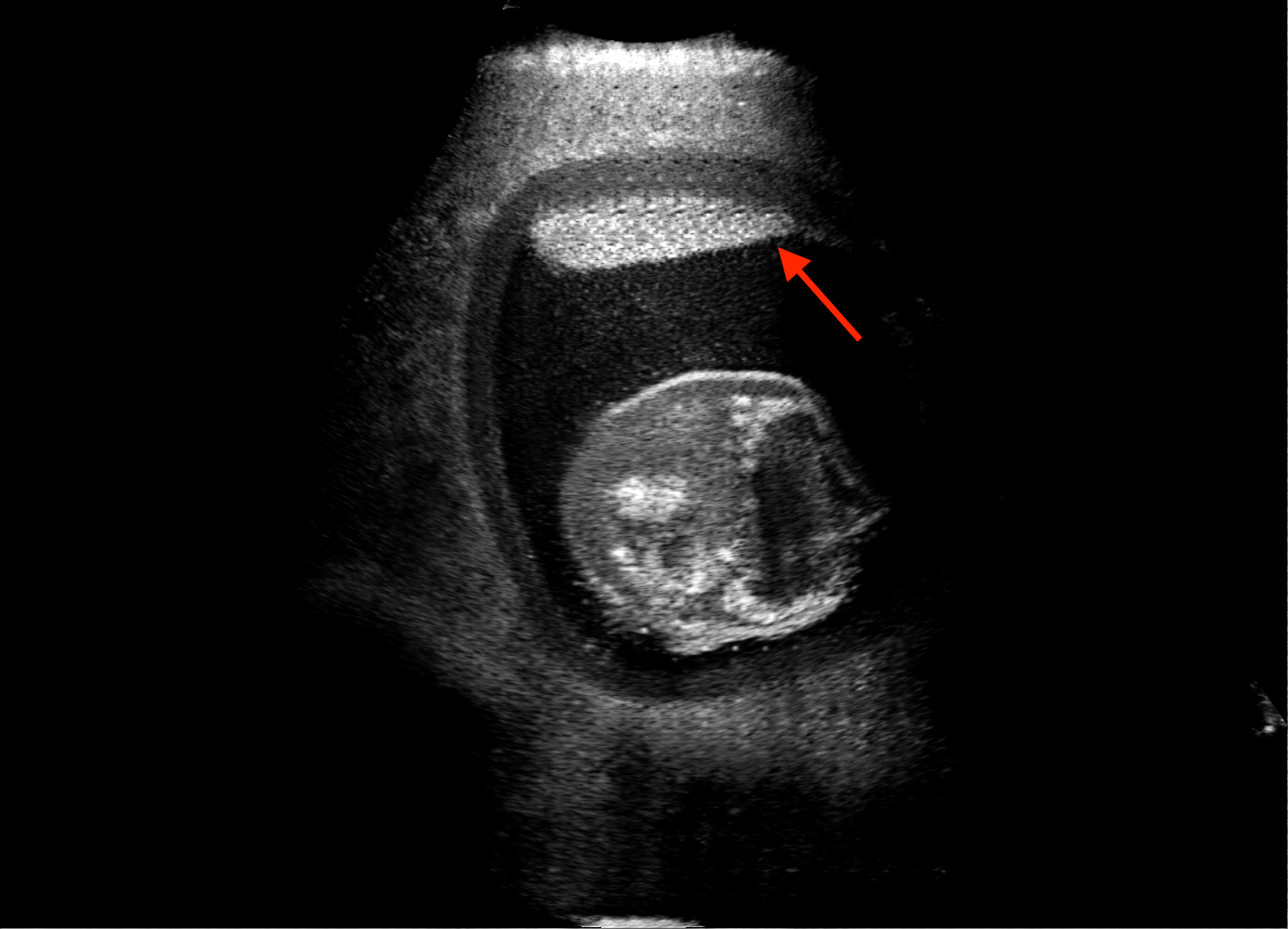}
\vspace{10pt}
\includegraphics[width=.22\textwidth]{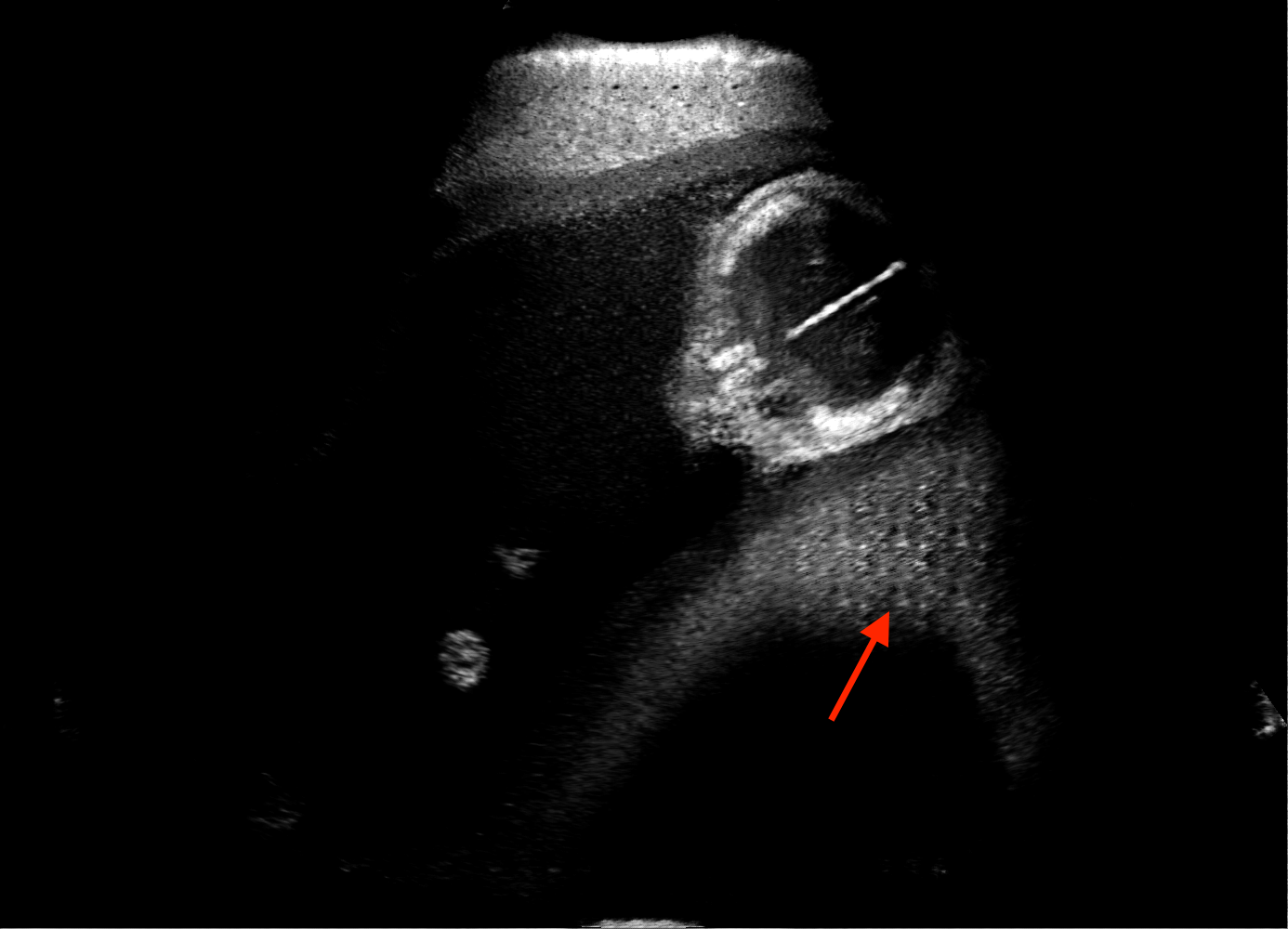}
\includegraphics[width=.22\textwidth]{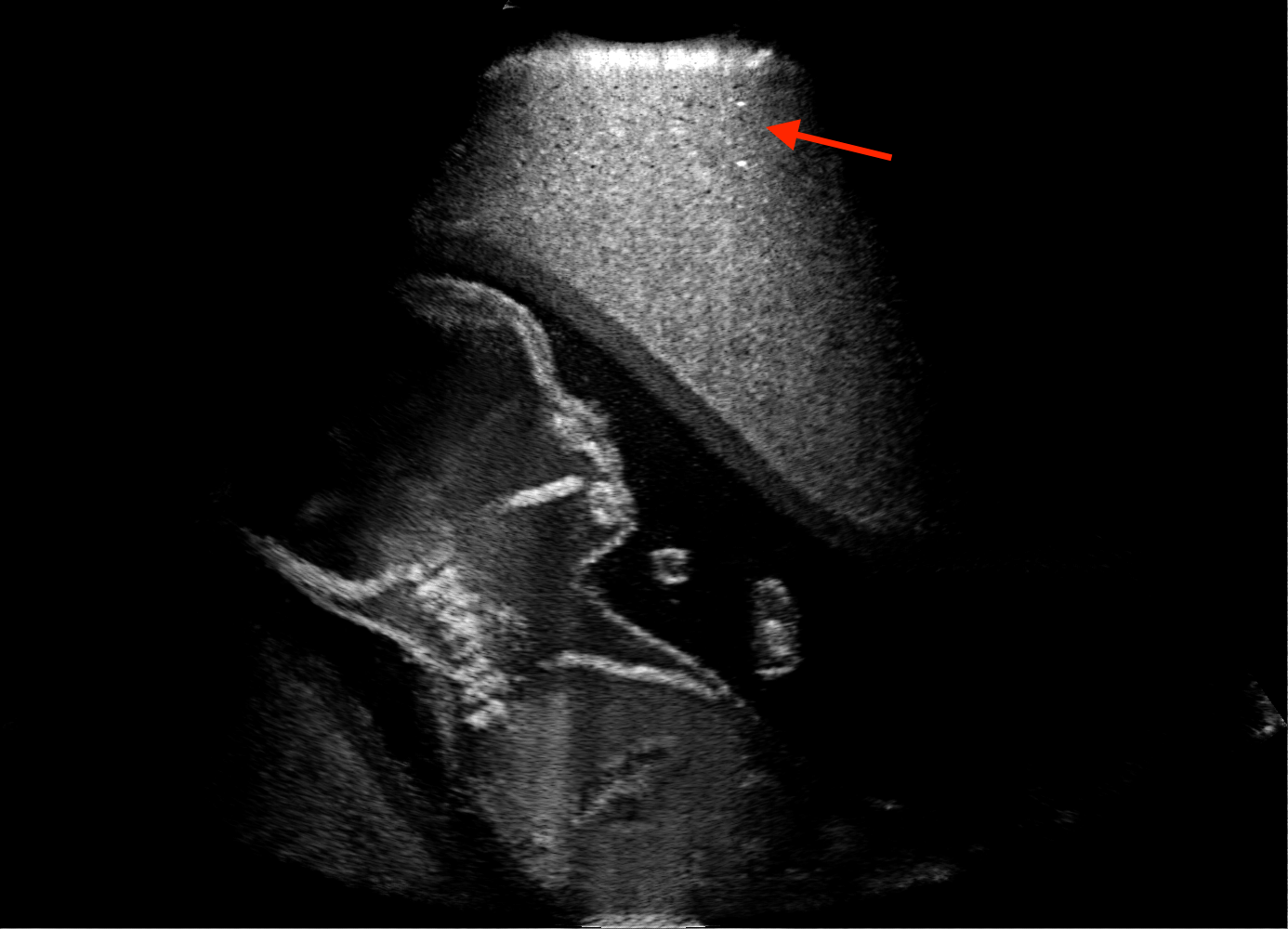}
\includegraphics[width=.22\textwidth]{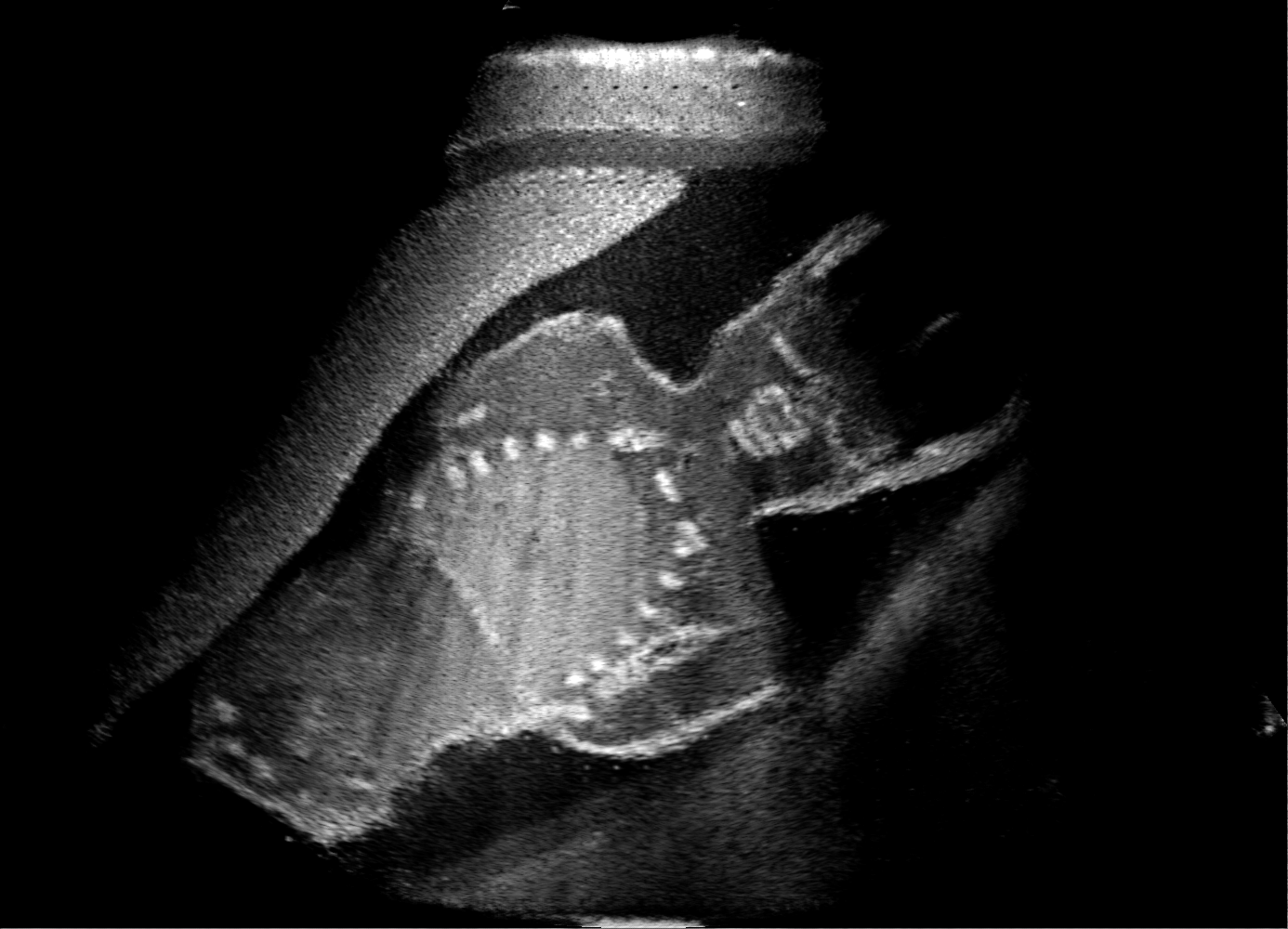}}
\caption{Inference on full field-of-view (FoV) images.
Segmentation and integral attenuation maps are shown, respectively, at the top left and bottom right corners of the simulated B-mode images.}
\label{fig:qual_results_hr}
\end{figure*}

\paragraph{Qualitative evaluation.} 
Fig.~\ref{fig:qual_results_hr} depicts the qualitative results for all the models mentioned above, with arrows pointing at structures relevant to discussion points below.
The visual results of the ablated variants of SA2H show substantial quality degradation compared to the full SA2H model, demonstrating the importance of each proposed architectural contribution.
Given only segmentation map in the network input, SA2H-att fails to generate acoustic shadows, e.g.\ those cast by the ribs.
Detailed structures such as the cervical vertebrae are blurred out in the SA2H-concat results, which also contain hallucinated structures mainly due to insufficient preservation of input information along the encoding-decoding path.
With SA2H-conv, checkerboard artefacts are observed due to the lack of proposed additional stride-1 convolutional layers.
SA2H-noise without any explicit noise input is seen to be sub-optimal at generating textural details.
The baseline method NSA2H fails to preserve anatomical structures and acoustic shadows in all cases, while the simulated textures also show significant artefacts such as checkerboard patterns.
Realism of different simulation aspects may become relevant given different clinical applications and scenarios. For instance, improved structural preservation, e.g. with the hyperechoic bony structures such as the skull and the ribs, of the final model over its ablated variants and NSA2H may prove relevant in fetal head measurements, while the textural improvements facilitating screening fetal organ maturity, e.g. lungs. Compared to the silver-standard model
LSA2H with a low-quality rendered image as additional input, SA2H is seen to be on par in structural preservation. Note that shadowing on homogenous regions (e.g. the rib shadowing on the homogenous lung region on the 4th column of Fig.~\ref{fig:qual_results_hr}) with our proposed method SA2H is represented more faithfully compared to LSA2H, whereas shadows on structurally complex regions (e.g. the skull shadowing around the heart and surrounding tissues on the 3rd column of Fig.~\ref{fig:qual_results_hr}) are suboptimal with our SA2H. Therefore, one may have to evaluate our method given particular simulation tasks, e.g. its clinical validity for fetal heart exams. However, even with low quality rendered images, LSA2H leads to artificial enhancements of intensities, lack of acoustic shadows, and low-quality textures especially near the probe, for which SA2H yields satisfactory results as illustrated in Fig.~\ref{fig:qual_results_hr}.

\paragraph{Quantitative evaluation.}
Metrics for assessment of US image quality and realism is an open research topic. Pixel-wise difference and SNR metrics assuming paired ground-truth images are often suboptimal and potentially inconclusive in judging US image realism, given the noisy speckle appearance and the inherent features and artifacts of B-mode images.
To assess such local image matching, we herein also utilize a sliding path-based histogram comparison metric.
Additionally, we utilize a typical metric for generative models that quantify distributions given features of a pretrained visual model.
Accorindingly, we utilize the following complementary quantiative metrics:

\begin{itemize}
\item[$\bullet$] {\bf Peak signal-to-noise ratio (PSNR)} is computed as $\textrm{PSNR} = 10\log_{10}(\frac{255}{\textrm{MSE}})$ with mean squared error MSE between two images.

\item[$\bullet$] {\bf Mean absolute error (MAE)} measures pixel-wise difference between two images.
High MAE value may indicate large intensity shift or structural mismatch.
However, it cannot provide information about texture difference.

\item[$\bullet$] {\bf p$\chi^2$}: We use $\chi^2$ distance to measure the difference in image histogram statistics, which are commonly used for tissue characterization~\cite{mailloux1985local,shankar1993use}.
This metric indicates potential mismatch in tissue speckle patterns, which affects image histogram statistics.
$\chi^2$ distance computes the difference between histograms $h_A$ and $h_B$ as:
$\chi^2 (h_A || h_{B}) = \frac{1}{2} \sum_{l=1..d} \frac{(h_A[l]-h_B[l])^2}{h_{A}[l]+h_{B}[l]}$. 
The number of histogram bins $d$ is set to $50$.
We compute this metric locally within non-overlapping sliding patches with a size of $20\times20$ to capture local textural information of US speckle patterns.
Herein we compute the root mean squared error, referred to \emph{patch $\chi^2$} (p$\chi^2$).

\item[$\bullet$] {\bf Fr\'{e}chet Inception Distance (FID)~\cite{heusel2017gans}} quantifies the similarity of generated samples to real samples by computing the distance between the feature vectors of the classification network Inception v3.
The feature vectors are fitted to multivariate Gaussian and the difference is computed using Fr\'{e}chet distance.
FID score is a widely used metric for assessing GAN performance.
Lower FID indicates better image quality.
We compute FID on $512\times512$ sized center crops of generated full field-of-view images, which are further divided into four sub-crops of $299\times299$, to increase the number of samples and match Inception v3 input size.
\end{itemize}

\begin{figure*}
\centering
\includegraphics[width=\textwidth]{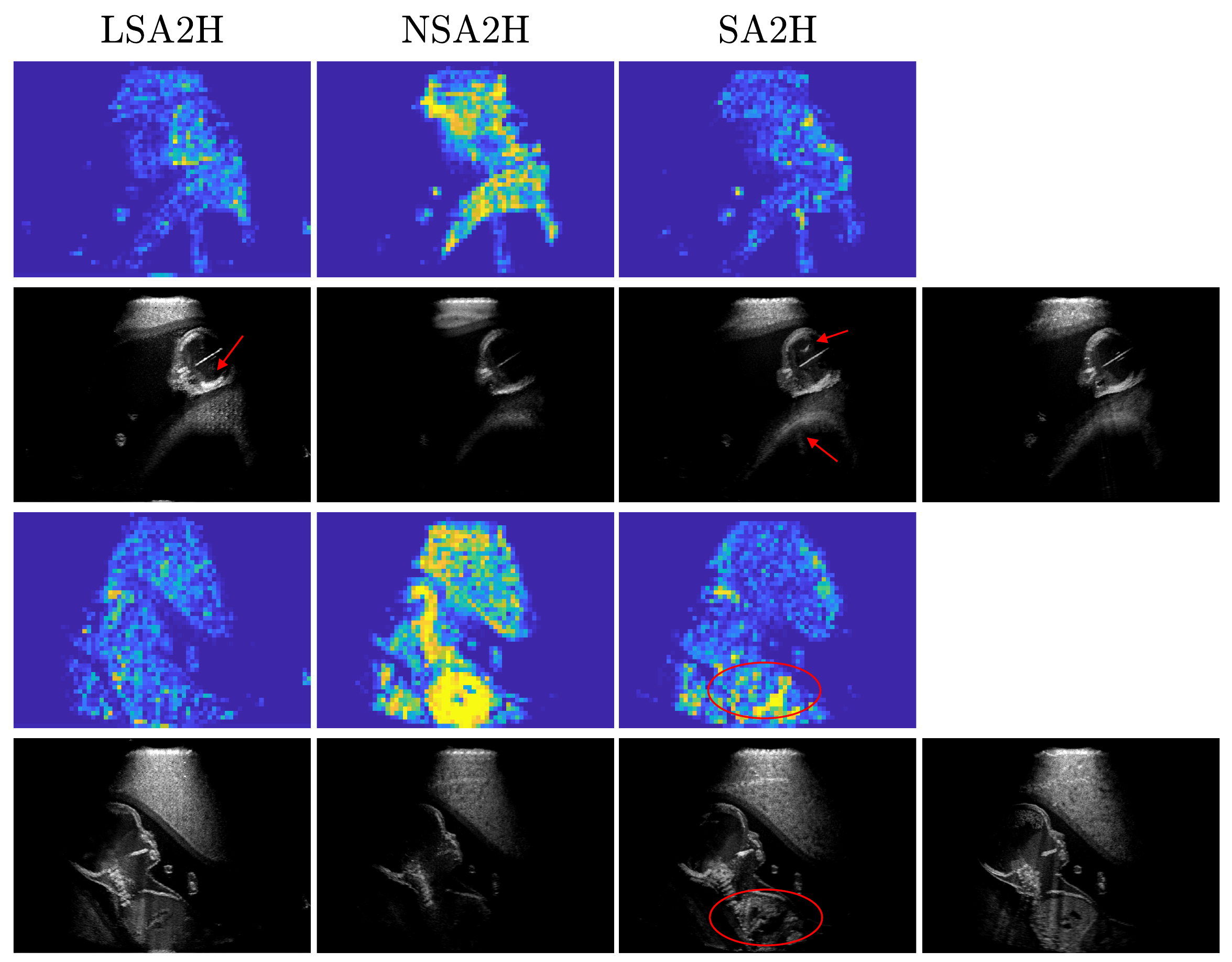}
\caption{Spatial p$\chi^2$ error map for the examples shown in Fig.~\ref{fig:qual_results_hr}. Error map is displayed within the range [0,1]. The corresponding ultrasound image is shown below the map. Target image is shown in the right most column.}
\label{fig:errmap}
\end{figure*}

For the interpretation of the local errors, sample spatial p$\chi^2$ error maps are depicted in Fig.~\ref{fig:errmap} for LSA2H, NSA2H, and SA2H for the middle two examples shown in Fig.~\ref{fig:qual_results_hr}.
Both images generated by NSA2H have a lot of missing structures and accordingly have high error almost all over the map.
Artificial skull enhancement with LSA2H is seem to evoke large p$\chi^2$ error, as shown in the corresponding error map, whereas the bright spots in the error map of SA2H reflect some hallucinated shadows and structure in the brain.
All of above mentioned regions of interest are marked by red arrows.
In the bottom example, SA2H fails to generate faithful content at the bottom region marked by red circles, which is well indicated by the error map as well.
\begin{table}
\setlength{\tabcolsep}{2pt}
\caption{Quantitative results. Bold number indicates the best performance. Mean and standard deviation are reported for PSNR, MAE and p$\chi^2$. Network capacity is given as number of trainable parameters.}
\centering
   \begin{tabular}{l|rr|rr|rr|r|r}
      & \multicolumn{2}{c|}{PSNR}& \multicolumn{2}{c|}{MAE} &
      \multicolumn{2}{c|}{$\textrm{p}\chi^2$ ($\times 10^{-2}$)}  &\multicolumn{1}{c|}{FID} &\multicolumn{1}{c}{\#params} \\
       \rowcolor{Gray}
 & mean & \quad std & mean & \quad std & mean & std & & \\
\hline
LSA2H & 27.38 & 0.52 &\bf{6.21} &1.56 & \bf{13.61} & 1.39  &64.42 &57.2M\\
NSA2H  & 25.59 & 1.81 &8.22 &2.77 & 31.39 & 5.51 &67.28 &57.2M\\
\hline
SA2H-att & 25.04 & 1.61 &8.75 &2.75 & 29.40 & 5.07  &92.49 &14.4M\\
SA2H-concat & 26.23 & 1.22 &9.08 &2.93 & 29.36 & 5.87  &76.40 &14.4M\\
SA2H-conv &24.85  &1.31 &8.87 &2.55  &32.70  &4.33  &93.18 &16.2M\\
SA2H-noise & 26.45 & 1.34 &8.14 &2.66 & 25.95 & 4.64  &97.85 &14.5M\\
\hline
SA2H  & \bf{28.19} & 1.23 &6.37 &2.08 &18.60 &5.23  &\bf{32.34} &14.5M
\end{tabular}
\label{tab:quant_results}
\end{table}

\begin{figure*}
\centering
\includegraphics[width=0.48\textwidth]{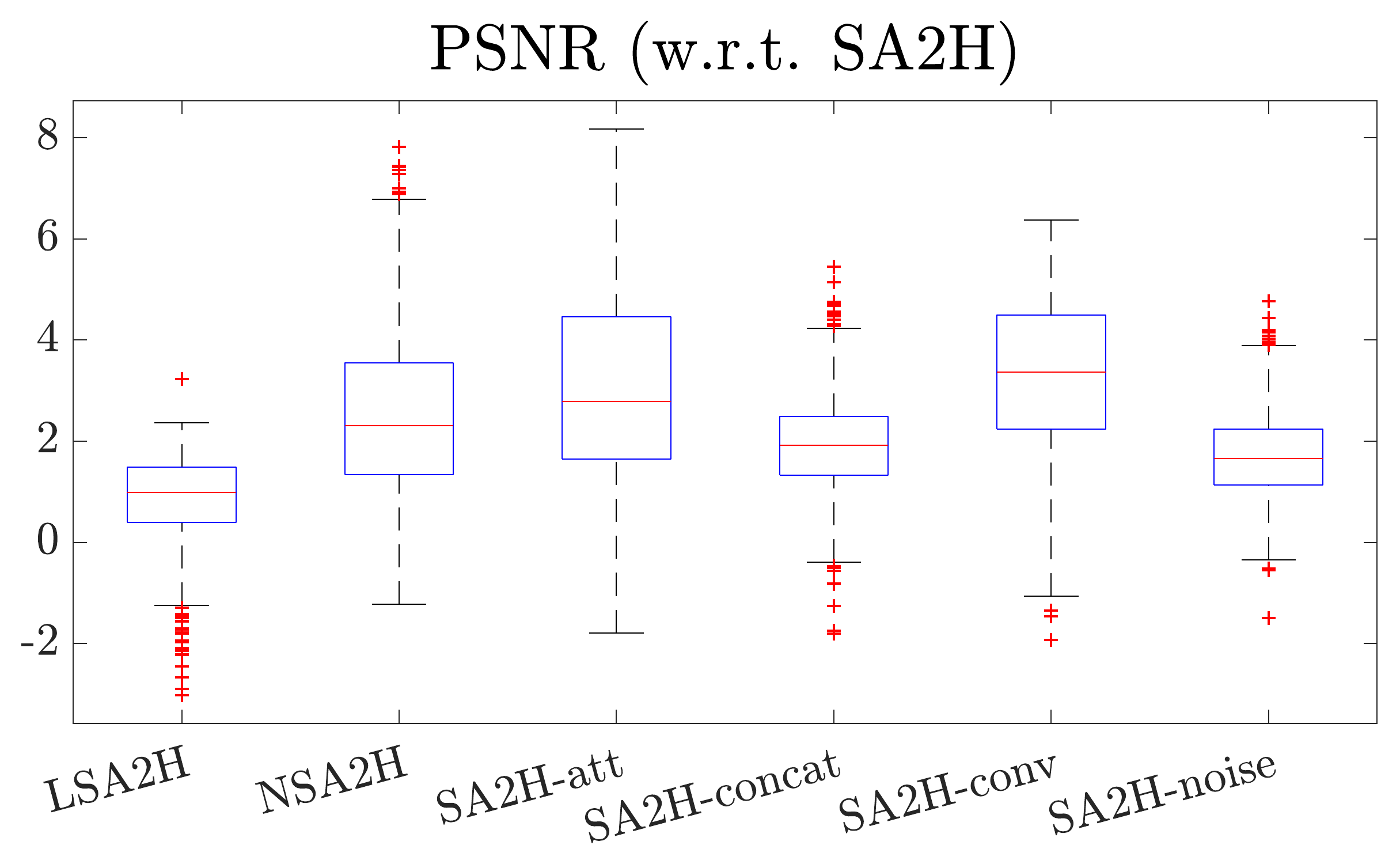}
\includegraphics[width=0.48\textwidth]{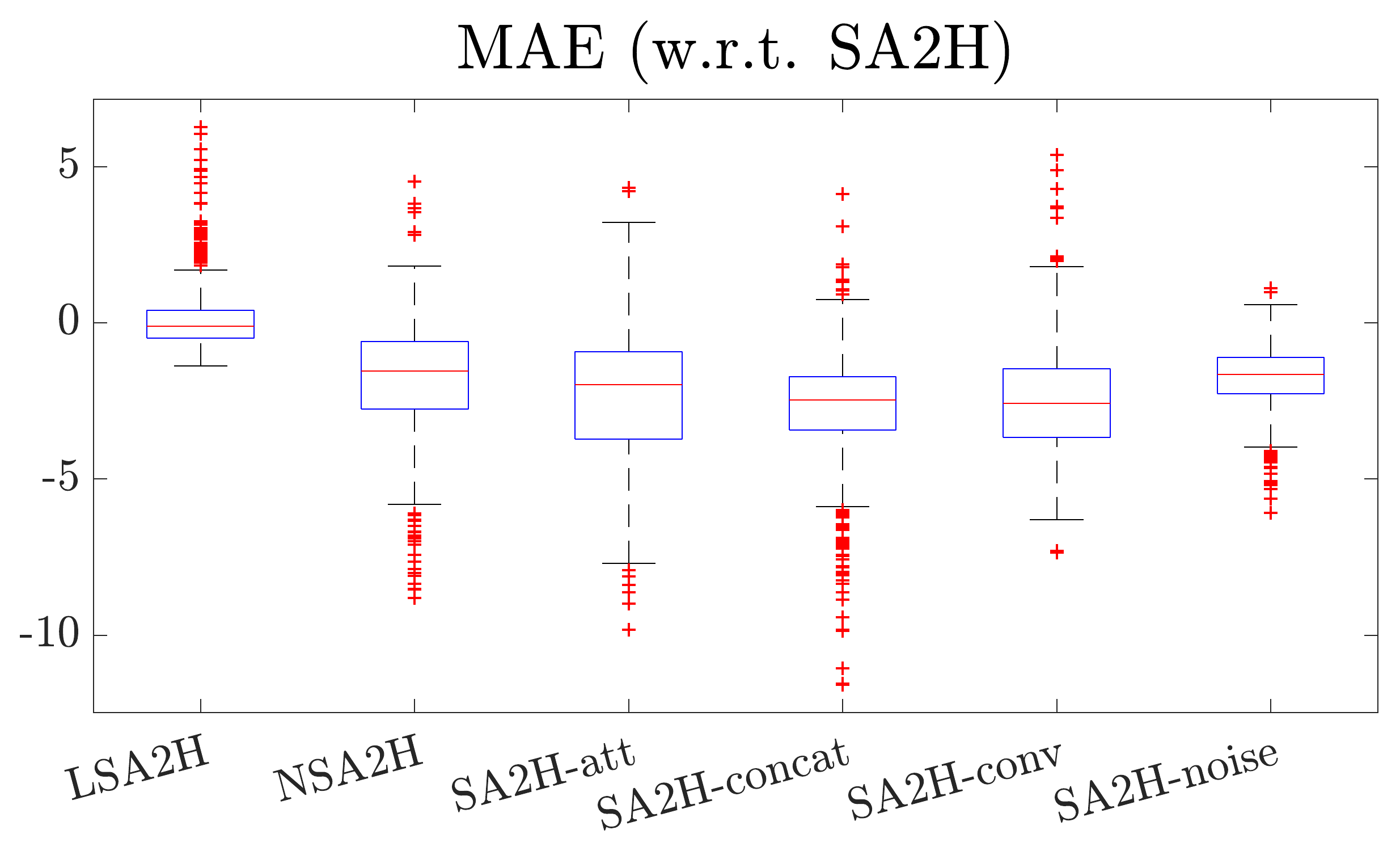}
\includegraphics[width=0.48\textwidth]{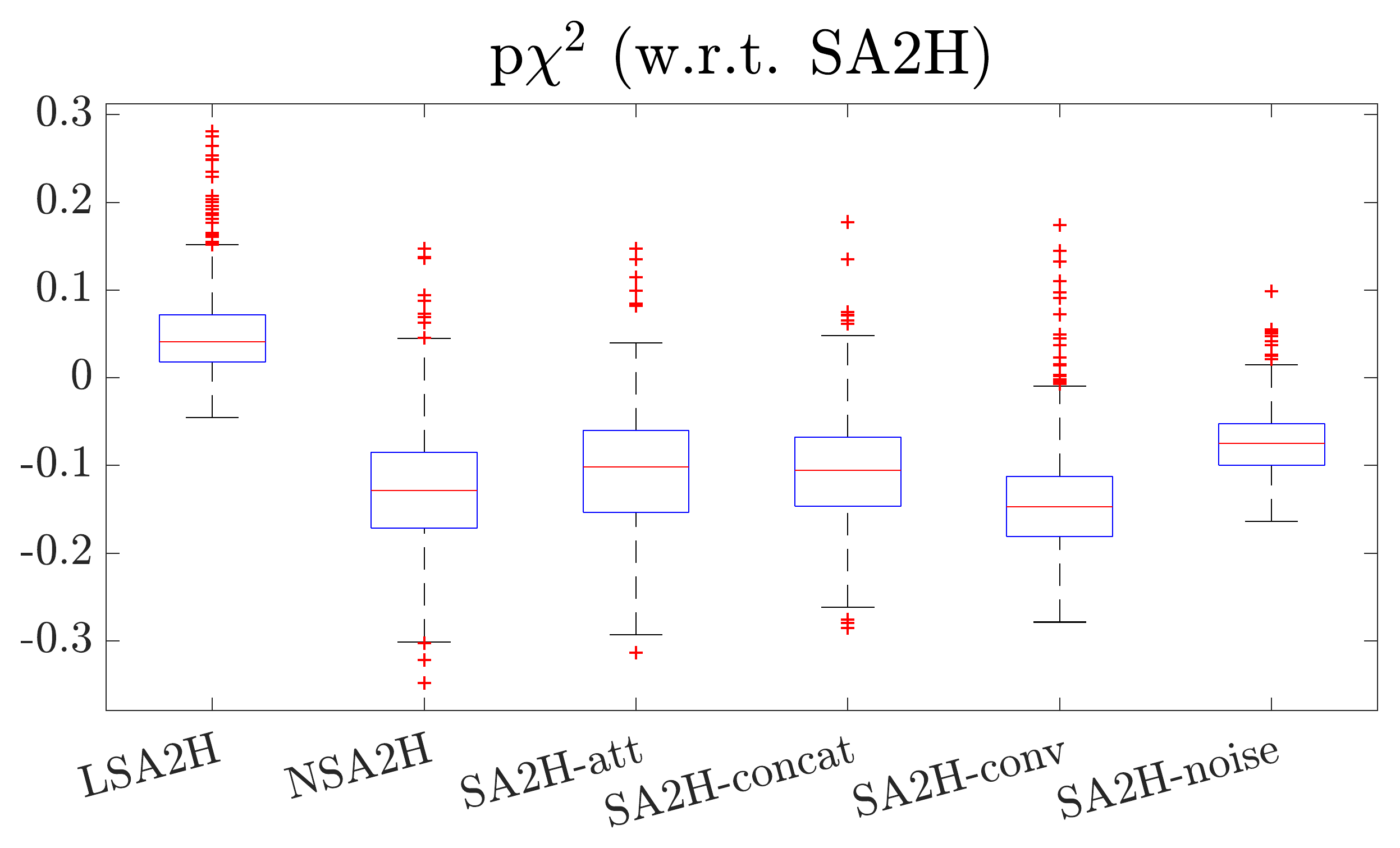}
\caption{Box plot of paired difference between SA2H and its ablated variants, LSA2H and NSA2H. Note that higher PSNR, lower MAE and $\textrm{p}\chi^2$ indicates better performance. For FID, no image-wise metric can be computed.}
\label{fig:boxplot}
\end{figure*}

The quantitative results are summarized in Tab~\ref{tab:quant_results}.
The effectiveness of all the proposed architectural improvements is well demonstrated by the significant performance drop of the ablated variants in all the metrics, which also corroborate with our qualitative results shown above.
The proposed model SA2H has achieves an improvement of over $50\%$ in FID and $40\%$ in $\textrm{p}\chi^2$ over the baseline NSA2H, indicating a significantly higher fidelity in generated images using the proposed method.
Surprisingly, for PSNR and FID metrics, SA2H outperforms the LSA2H model, which has extra ultrasound information provided as a low-quality image input.
In Fig.~\ref{fig:boxplot} we show the distribution of paired differences in PSNR, MAE, and p$\chi^2$ between each tested model and the proposed SA2H one.
As seen, for all ablated variants and the baseline NSA2H, our proposed method provides a significant improvement for all metrics.

The presented method is aimed to replace the computationally expensive online rendering process, such that image quality is not restricted by the interactivity constraint of a simulation complexity and runtime. 
Using training images from a wide range of positions and orientations on a given abdominal model, our generator learns to simulate B-mode images from arbitrary slices of this model, as assessed with unseen image locations in our evaluation.
For other anatomical models or locations, one can retrain the network with their corresponding simulated images and cross-sections.

Note that paired segmentation/attenuation maps and US images are required to train our network. 
The choice of training data is therefore restricted to synthetic images. Hence, the realism of the generated images by our proposed approach cannot go beyond the realism of the underlying simulation. To incorporate real ultrasound images, one may utilize  unpaired deep learning models, e.g. cycleGAN~\cite{zhu2017unpaired,vitale2019improving} , as a potential future research direction.

Our work similarly to many other GAN-based techniques for US simulation~\cite{hu2017freehand,tom2018simulating,vitale2019improving} deal with single image translation or generation. 
However, for sonographer training purposes, interactive and dynamic simulators are required to generate temporally consistent image sequences. 
However, without explicitly enforcing temporal consistency, these methods are not guaranteed to generate images free from temporal artifacts, which is a potential research direction for future work.

\section{Conclusions}
\label{sec:3}
In this work, we demonstrate a GAN-based framework with a local feature preserving generator architecture for learning ultrasound rendering from cross-sectional segmentation and subsequent integral attenuation maps, the latter of which greatly facilitates the generation of directional acoustic shadows.
Combining with texture-friendly decoder blocks and a proposed noise feeding strategy, the presented network improves quality of translated images in terms of structure and texture compared to the state-of-the-art and any baseline method with an improvement of over $50\%$ in FID score.
An extensive ablation study has been carried out to demonstrate the effectiveness of each proposed architectural contribution.
Compared to an earlier approach of simulated ultrasound generation using image translation from low-quality rendered images, the current approach does not require sophisticated online rendering step, while still able to generate ultrasound images with good anatomical structural correspondence and superior texture appearance compared to that state-of-the-art.
For evaluation, besides traditional metrics we also propose a local histogram statistics based metric, while demonstrating on examples how it captures visually-perceived local differences between images, which is a typically challenging task due to the inherent speckle noise in ultrasound images.

Learning rendering with GANs warrants sophisticated rendering for ultrasound simulation to be carried out on consumer-hardware.
The current rendering simulation settings lead to a frame time of 75 ms, a low-end of visually-acceptable US interactivity. 
Our GAN requires a constant computation time of 40 ms per frame using TensorRT, hence nearly doubling the frame rate.
Moreover, pre-trained network can further be efficiently transferred and implemented on low-end devices, such as with FPGAs~\cite{guo2017survey}.
With a wider availability of simulated training of ultrasound with realistic imagery, future sonographers can be trained more effectively and also for rare-cases both for diagnostic and interventional imaging applications.

\bibliographystyle{IEEEtran}
\bibliography{image_translation_us}
\end{document}